\documentclass[a4paper,fleqn,usenatbib]{mnras}

\usepackage{newtxtext,newtxmath}
\usepackage[T1]{fontenc}
\usepackage{ae,aecompl}

\usepackage{graphicx}	% Including figure files
\usepackage{amsmath}	% Advanced maths commands
\usepackage{amssymb}	% Extra maths symbols

\title[Spectral analysis of four `hypervariable' AGN: a microneedle in the haystack?]{Spectral analysis of four `hypervariable' AGN: a microneedle in the haystack?}

\author[Bruce et al.]{
A. Bruce,$^{1}$\thanks{E-mail: alb@roe.ac.uk (AB); al@roe.ac.uk (AL)}
A. Lawrence,$^{1\color{blue}\star}$
C. MacLeod,$^{2}$
M. Elvis,$^{2}$
M. J. Ward,$^{3}$
J. S. Collinson,$^{3}$
\newauthor
S. Gezari,$^{4}$
P. J. Marshall,$^{5}$
M. C. Lam,$^{1}$
R. Kotak,$^{6}$
C. Inserra,$^{6}$
J. Polshaw,$^{6}$
N. Kaiser,$^{7}$
\newauthor
R-P. Kudritzki,$^{7}$
E. A. Magnier$^{7}$ and
C. Waters$^{7}$
\\
$^{1}$Institute for Astronomy, SUPA (Scottish Universities Physics Alliance), University of Edinburgh, Royal Observatory, Blackford Hill, Edinburgh EH9 3HJ, Edinburgh, UK\\
$^{2}$Harvard-Smithsonian Center for Astrophysics, 60 Garden St., Cambridge, MA 02138, USA\\
$^{3}$Department of Physics, University of Durham, South Road, Durham, DH1 3LE, UK\\
$^{4}$Department of Astronomy, University of Maryland, College Park, MD  20742-2421, USA\\
$^{5}$Kavli Institute for Particle Astrophysics and Cosmology, PO Box 20450, MS 29, Stanford, CA 94309, USA\\
$^{6}$Astrophysics Research Centre, School of Mathematics and Physics, Queens University Belfast, Belfast BT7 1NN, UK\\
$^{7}$Institute for Astronomy, University of Hawaii, 2680 Woodlawn Drive, Honolulu, HI 96822, USA\\
}

\date{Accepted 2017 January 18. Received 2016 December 23; in original form 2016 August 22}

\pubyear{2017}

\begin{document}
\label{firstpage}
\pagerange{\pageref{firstpage}--\pageref{lastpage}}
\maketitle

% Abstract of the paper

\begin{abstract}
We analyse four extreme active galactic nuclei (AGN) transients to explore the possibility that they are caused by rare, high-amplitude microlensing events. These previously unknown type-I AGN are located in the redshift range 0.6--1.1 and show changes of > 1.5 mag in the $g$ band on a time-scale of $\sim$years. Multi-epoch optical spectroscopy, from the William Herschel Telescope, shows clear differential variability in the broad line fluxes with respect to the continuum changes and also evolution in the line profiles. In two cases, a simple point-source, point-lens microlensing model provides an excellent match to the long-term variability seen in these objects. For both models, the parameter constraints are consistent with the microlensing being due to an intervening stellar mass object but as yet there is no confirmation of the presence of an intervening galaxy. The models predict a peak amplification of 10.3/13.5 and an Einstein time-scale of 7.5/10.8 yr, respectively. In one case, the data also allow constraints on the size of the \ion{C}{iii]} emitting region, with some simplifying assumptions, to be $\sim1.0\textup{--}6.5$ light-days and a lower limit on the size of the \ion{Mg}{ii} emitting region to be $>9$ light-days (half-light radii). This \ion{C}{iii]} radius is perhaps surprisingly small. In the remaining two objects, there is spectroscopic evidence for an intervening absorber but the extra structure seen in the light curves requires a more complex lensing scenario to adequately explain.
\end{abstract}

% Select between one and six entries from the list of approved keywords.
% Don't make up new ones.

%%%%%%%%%%%%%%%%%%%%%%%%%%%%%%%%%%%%%%%%%%%%%%%%%%

%%%%%%%%%%%%%%%%% BODY OF PAPER %%%%%%%%%%%%%%%%%%

\section{Introduction}

The identification of a class of active galactic nuclei (AGN) transients that are smoothly evolving, by factors of several, on year-long time-scales (\citealt{Lawrence2012,Lawrence2016}, hereafter L16) has raised a number of interesting questions regarding the underlying cause. These `hypervariable' AGN may simply be at the extreme end of the tail of typical AGN variability \citep{MacLeod2010,MacLeod2012} or, perhaps more interestingly, there may be an extrinsic cause. Plausible mechanisms for these transients include the following: accretion events/instabilities; tidal disruption events (TDEs) or extinction events. A fourth possibility, and the focus for this particular work, is that some of these AGN transients are actually rare, high-amplitude microlensing events. If this is the case, through analysis of multi-epoch spectroscopy and simple lensing models, we have the potential to uncover valuable information on the innermost regions of these enigmatic objects.

Observational studies of the microlensing seen in multiply imaged AGN are well established and provide the means with which one can ascertain properties intrinsic to those objects \citep{Irwin1989,Eigenbrod2008,Morgan2010,Blackburne2011,Mosquera2011,Jimenez-Vicente2012,Jimenez-Vicente2014,Sluse2012,MacLeod2015}. This involves a monitoring of the AGN components in order to ascertain both the level of variability intrinsic to the AGN and that due to microlensing, typically a low-level `flickering', in one or more of these components. In contrast, the AGN in this paper exhibit no clear sign of strong lensing effects and are varying significantly and smoothly (relative to typical AGN behaviour). If a rare, high-amplitude microlensing event is the cause of this variability, this places these previously unknown AGN in a different regime to their multiply imaged counterparts. 

In this paper, four promising AGN transients are selected as candidates for an exploration of the microlensing scenario and its consequences. That is, the hypothesis that an intervening stellar mass object is responsible for the bulk variability seen in these objects. It should be noted that microlensing events will not explain all hypervariable AGN activity; rather, they likely describe a subset of this interesting population. In the interest of brevity, this paper will focus entirely on the microlensing scenario but interested readers should refer to L16 and references therein for more discussion on the alternative variability scenarios.

\defcitealias{Planck2014}{Planck Collaboration XVI 2014}

Section \ref{sec:Observations} describes our sample selection, observational data and reduction pipeline. Spectroscopic results are detailed in Section \ref{sec:SpecResults}. Section \ref{sec:MicroModels} outlines the procedures used for exploring different lensing models with results detailed in Section \ref{sec:MicroResults}. In Section \ref{sec:Discussion}, the implications for future work and the advantages/disadvantages over existing microlensing studies involving multiply imaged AGN are discussed. Cosmological calculations in this paper make use of Planck13 values (\citetalias{Planck2014}; $H_0=67.8, \Omega_\Lambda=0.693$).

\section{Observations}
\label{sec:Observations}

\subsection{Target selection}
\subsubsection{Parent sample}
The four objects in this paper are selected from a larger sample of highly variable AGN, discovered as part of a wider transient search initially designed to look for TDEs around quiescent black holes. A significant fraction of candidate events varied on longer time-scales than that expected for TDEs or supernovae and were subsequently revealed to be AGN. The transients were identified by looking for changes of greater than 1.5 mag in the Panoramic Survey Telescope and Rapid Response System (Pan-STARRS) 1 $3\pi$ Steradian Survey data when compared with the Sloan Digital Sky Survey (SDSS) footprint from around a decade earlier. In addition, the transient had to be located within 0.5 arcsec of an object classified as a `galaxy' in the DR7 catalogue. Approximately two-thirds of the flagged objects have been spectroscopically confirmed as type-I AGN and it is these that make up the larger `hypervariable' AGN sample, currently 63 objects. For further details, see L16.

\subsubsection{Selection criteria for this paper}
\label{sec:Criteria}
To explore the microlensing hypothesis, objects were selected from the hypervariable AGN sample based on the following criteria. First, that the photometry displays signs of smooth evolution on long time-scales with a change in magnitude of $\Delta g>0.5$ mag over the period of spectral observations. Secondly, that there are a minimum of two spectral observations separated by at least one year (observed frame).  The former favours microlensing events over intrinsic variability and the latter increases the chance of observing spectroscopic trends, allowing further testing of the microlensing hypothesis.

Additional factors that favour a microlensing scenario are summarized in Table \ref{tab:criteria}. Here, the presence of a near-symmetric single- or double-peaked structure in the light curve, or evidence for an intervening galaxy, and hence lens repository, are noted. At least one of these additional factors was required in the target selection process leaving a total of four microlensing candidates for consideration. Any evidence for an intervening galaxy will be discussed as part of the spectroscopic results for that target (Section \ref{sec:SpecResults}).

\begin{table}
	\centering
	\caption{Summary of the target selection criteria noted in Section \ref{sec:Criteria}}
	\label{tab:criteria}
	\begin{tabular}{llccc}
		\hline
		Short ID & SDSS ID & Single-  & Double- & Int-\\
				 &		   & peak?    & peak?   & gal?\\
		\hline
		J084305 & J084305.54+550351.3 & Yes & No  & No \\
		J094511 & J094511.08+174544.7 & Yes & No  & No \\
		J142232 & J103837.08+021119.6 & No  & Yes & Yes \\
		J103837 & J142232.45+014026.8 & No  & No  & Yes \\
		\hline
	\end{tabular}
\end{table}

\subsection{Photometry}

\subsubsection{SDSS data}

SDSS data used in this paper are DR9 \citep{Ahn2012} \textit{cmodel} $g$-band magnitudes. These should provide the most reliable magnitude estimate for an extended object in a single band and should also agree with the Point Spread Function (PSF) magnitudes for stars. Not all bands are best fit with extended models for these objects, as might be expected with the presence of an AGN component. Also, from DR7 to DR9, J103837 has had the classification changed to stellar. The least reliable data is that for J142232, which, in addition to being faint, is flagged as having been affected by a cosmic ray hit.

\subsubsection{Pan-STARRS data}

The Pan-STARRS 1 3$\pi$ $g_{\rm P1}$ data used in this paper are from the PV1.2 data release \citep{Schlafly2012,Tonry2012,Magnier2013} and magnitudes have been calculated from the PSF fluxes and associated zero-points.

\subsubsection{Liverpool Telescope data}

The Liverpool Telescope (LT) provides finer sampling of each transient in the AGN sample and was also instrumental in the classification of the Pan-STARRS transients. The LT is a fully robotic 2.0m telescope situated on La Palma and operated by Liverpool John Moore's University \citep{Steele2004}. The transients were initially monitored every few days in $u_{\rm LT}$, $g_{\rm LT}$ and $r_{\rm LT}$ to determine how fast they were evolving and then roughly every few weeks. Scheduling and weather constraints sometimes meant that there was not always the desired cadence on some targets. The observing programme with the LT is still ongoing. Only the $g_{\rm LT}$ data is utilized in this paper.

\subsubsection{CRTS data}

In addition to the SDSS and LT photometry, there are data from the Catalina Real-Time Transient Survey (CRTS;  \citealt{Drake2009}). This survey makes use of three different telescopes and allows us to recover pre-Pan-STARRS era light curves for some of the objects. A note of caution is that this survey uses clear filters calibrated to a $V$-band zero-point, so colour effects may be significant. An approximate magnitude offset has been applied so that the light curve appears consistent with the other data sets for each target and the data have also been seasonally averaged for clarity. Due to the uncertainty regarding colour effects, the CRTS data is not used in the modelling process.

\subsubsection{Filter approximation}

For the purposes of this paper, we assume that the differences between the $g_{\rm SDSS}$, $g_{\rm P1}$ and $g_{\rm LT}$ filters can be neglected and designate magnitudes as simply $g$. This should be reasonable given that we expect the photometric uncertainty to be dominated by intrinsic AGN variability, typically ~0.1 mag or greater. For the microlensing models, magnitude-to-flux conversions are performed assuming an effective wavelength of 4770\,\AA.

\subsection{Spectroscopy}
\subsubsection{WHT data}

The majority of the spectral observations were performed on the 4.2m William Herschel Telescope (WHT), La Palma, using the Intermediate dispersion Spectrograph and Imaging System (ISIS) long-slit spectrograph. The 5300 dichroic was used along with the R158B/R300B grating in the red/blue arms, respectively, along with the GG495 order sorting filter in the red arm. Typically 2$\times$ binning in the spatial direction was used to improve the signal-to-noise ratio (SNR) along with a narrow CCD window to reduce disk usage and readout times. This set-up gives a spectral resolution of $R\sim$1500 at 5200\,\AA\, in the blue and $R\sim$1000 at 7200\,\AA\, in the red for a slit width of 1 arcsec and total coverage $\sim3100\textup{--}10600$\,\AA.

Typically calibration images were taken at the start of each night including bias frames, lamp flats and CuNe/Ar arc lamp images. Spectroscopic standard stars were imaged at $\sim$2h intervals throughout the night though this cadence was not always possible. Observations were carried out at the parallactic angle and further observational details for each target in this paper are given in Table \ref{tab:spec_obs}. Exposures were taken in $1800{\rm s}$ increments and the number of shots on target was adjusted based on the latest $g_{\rm LT}$ photometry.

\begin{table*}
	\caption{Details of the spectral observations carried out for the objects listed in this paper. The airmass values reflect the mean airmass over the duration of the observation(s). Magnitudes estimates are approximations based on the LT photometry. Seeing estimates reflect the range in measured full width at half maximum (FWHM) of the central region of the blue arm trace for each image.}
	\label{tab:spec_obs}
	\centering
	\begin{tabular}{rllllllll}
		\hline
		Target & Scope & Date & Exposures & $g_{\rm mag}$ & Slit & Seeing & Airmass & Notes\\
		\hline
		J084305--1 & WHT & 2013-02-09 & $1\times1800$\,s                & 20.0 & 2 arcsec   & 2.2 arcsec     & 1.74      & Dark, patchy cloud, variable seeing\\		
		       --2 & WHT & 2013-03-31 & $2\times1800$\,s                & 20.2 & 1 arcsec   & 1.1--1.2 arcsec & 1.12--1.14 & Moon 73\%(sep $122^\circ$), stable\\		
		       --3 & WHT & 2014-12-17 & $4\times1800$\,s                & 21.1 & 2 arcsec   & 2.0--2.7 arcsec & 1.51--1.24 & Dark, variable seeing\\		
		       --4 & MMT & 2015-03-10 & $2\times1800$\,s                & 21.3 & 1.5 arcsec & 0.8--1.1 arcsec & 1.1--1.09  & Moon 82\%(sep $96^\circ$), variable seeing\\		
		J094511--1 & WHT & 2013-05-15 & $2\times1800$\,s                & 20.7 & 1 arcsec   & 1.3 arcsec     & 1.18--1.28 & Moon 30\%(sep $24^\circ$), stable\\
		       --2 & WHT & 2014-02-07 & $4\times1800$\,s                & 21.1 & 1 arcsec   & 1.0--1.4 arcsec & 1.34--1.08 & Moon 63\%(sep $80^\circ$), light cloud\\
		       --3 & MMT & 2015-03-09 & $2\times1800$\,s                & 21.5 & 1.5 arcsec & 0.8--0.9 arcsec & 1.06--1.04 & Moon 89\%(sep $63^\circ$), variable seeing\\
		J142232--1 & WHT & 2013-02-11 & $2\times1800$\,s                & 20.0 & 1.5 arcsec & 1.8--2.0 arcsec & 1.24--1.18 & Dark, variable seeing\\
		       --2 & WHT & 2013-05-15 & $1\times1800$\,s                & 20.1 & 1 arcsec   & 1.8 arcsec     & 1.34      & Dark, stable\\
		       --3 & WHT & 2013-08-07 & $1\times1800$\,s                & 20.2 & 1 arcsec   & 1.1 arcsec     & 1.58      & Dark, stable\\
		       --4 & WHT & 2014-02-07 & $2\times1800$\,s                & 20.6 & 1 arcsec   & 1.1--1.2 arcsec & 1.14--1.13 & Dark, stable\\
		       --5 & WHT & 2014-07-24 & $4\times1800$\,s                & 21.0 & 1 arcsec   & 1.6--2.1 arcsec & 1.25--1.93 & Dark, stable\\
		          &     &          & $1\times900$\,s                 & &                 & &                          & \\
	  (unused) --6 & MMT & 2015-03-09 & $2\times1800$\,s                & 21.0 & 1 arcsec   & -- & 1.38--1.28 & Moon 89\%(sep $14^\circ$), variable, v. poor\\
		       --7 & WHT & 2015-04-23 & $4\times1800$\,s                & 21.0 & 1 arcsec   & 1.6--1.9 arcsec & 1.25--1.68 & Dark, stable\\
		J103837--1 & WHT & 2013-02-11 & $1\times1800$\,s                & 19.7 & 1.5 arcsec & 1.9 arcsec     & 1.13      & Dark, variable seeing\\
		       --2 & WHT & 2013-05-15 & $1\times1800$\,s                & 20.2 & 1 arcsec   & 1.3 arcsec     & 1.38      & Moon 30\%(sep $40^\circ$), stable\\
		       --3 & WHT & 2015-04-21 & $2\times1800$\,s                & 19.8 & 1 arcsec   & 0.8--0.9 arcsec & 1.13--1.12 & Dark, stable\\
		\hline
	\end{tabular}
\end{table*}

\subsubsection{MMT data}

A small number of observations were made with the blue arm spectrograph on the 6.5m MMT situated on Mount Hopkins, Arizona. Here, the $300\textrm{\,g\,mm}^{-1}$ grating was used with $2\times$ binning in the spatial direction. A filter wheel issue meant that no order-sorting filter was used. This may affect the third epoch for target J094511 and fourth epoch for J084305. Due to variable conditions and a nearby bright Moon, the sixth epoch for J142232 was very poor and will not be used in the analysis.

\subsubsection{Spectroscopic reduction pipeline}
\label{sec:Pipeline}

A reduction pipeline was created using custom \textsc{pyraf} scripts and standard techniques. After bias-subtraction and flat-fielding the cosmic rays were removed using the \textsc{lacos\_spec} script \citep{VanDokkum2001}. The spectra were then extracted and wavelength calibrated using the arcs obtained that night. In order to minimize problems with the calibration due to temperature variations or instrument flexure, an additional step was to offset each red/blue spectrum by a small amount, typically $\sim\pm0\textup{--}3\,$\AA, to ensure the prominent atmospheric oxygen line at 5577.338\,\AA\, had the correct wavelength in all observations. Flux calibration was performed using a single standard star and the mean extinction curve for the observatory. Where target observations were bracketed by standard star observations, the one least affected by transparency issues was used, i.e. the standard with the superior {\tt sensfunc} response. To better enable the combination of the red and blue data, the inner wing of each spectrum, where the response of the instrument declines steeply, was flux calibrated separately. The final step was to average all calibrated spectra and rebin to a linear wavelength scale.

Long-slit spectroscopy can suffer from transparency problems, which affect the absolute flux calibration. In order to minimize these effects, each target spectrum has been rescaled. For three of the targets, a smooth interpolation through the LT photometry was used to correct the spectra. In the case of J084305 and J094511, this was done by rescaling to the working microlensing model. For J142232, this was accomplished using a cubic-spline fit to the LT light curve. The fourth target, J103837, does not produce a satisfactory cubic-spline fit at all epochs due to a poorer LT cadence. Instead, the spectra for this target have been rescaled, so that the measured \ion{[O}{iii]}$_{5007}$ flux, tied to the third epoch, remains constant. This method will suffer if there are significant seeing changes, and it may even be possible that on these year-long time-scales, there will be some intrinsic narrow line variability \citep{Peterson2013}. Spectral fluxes were measured using an LT $g$-band transmission curve\footnote[1]{http://telescope.livjm.ac.uk/TelInst/Inst/IOO/}. The scaling factors applied are noted in Table \ref{tab:data}. Though the \ion{Mg}{ii} broad emission line is present within the $g$ band for these objects (excluding J142232), we have not attempted to account for this in the rescaling process so as to keep the data consistent with the photometry.

\subsubsection{Spectral fitting}
\label{sec:Spectral_fitting}

\begin{figure}
	\includegraphics[width=\columnwidth]{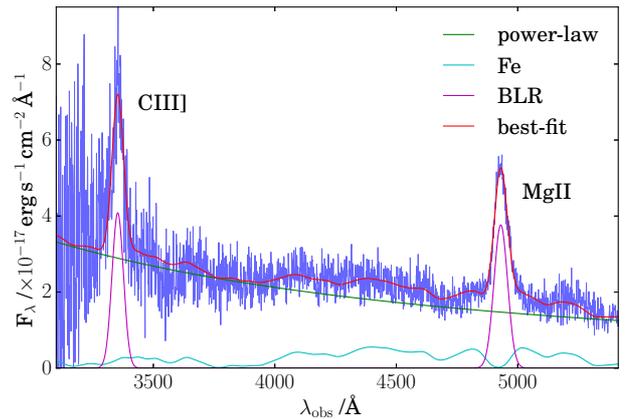}
    \caption{Example of the fitting process for target J094511. The components include the following: power-law (green), Gaussian line fits (magenta), convolved iron template (cyan) and overall best fit (red).}
    \label{fig:fit_example}
\end{figure}

For the spectral fitting process, a \textsc{python} package, \textsc{lmfit}, was used. This is a non-linear optimization and curve-fitting tool that builds on a Levenberg--Marquardt algorithm. It was used to fit a single Gaussian component to the emission lines and provide a power-law fit to approximate the local continuum. The relatively low SNR in the blue arm for some epochs ultimately ruled out the use of a multicomponent fit for most broad lines. The power law used in the fitting routine takes the form
\begin{equation}
	F_\lambda=A(\lambda/5100\,$\AA$)^\beta,
	\label{eq:powerlaw}
\end{equation}
where $A$ is the normalization and $\beta$ is the power-law slope. In addition to the above components, a template fit was used to estimate the Fe contribution. In the UV, the empirical template is that from \citet{Vestergaard2001}, and in the optical, where possible, that of \citet{Veron-Cetty2004}. The Fe template is convolved with a Gaussian in order to better approximate the true blended Fe emission. The width of the convolving Gaussian was set to match that of the broad \ion{Mg}{ii}/H\,$\beta$ components in the UV/optical, respectively. Before performing any fits the spectra are first corrected for Milky Way extinction using the A$_{\textrm V}$ values in Table \ref{tab:target_info} and the extinction law in the optical from \citet{Cardelli1989}. No attempt has been made to correct for host galaxy reddening at this stage. The most prominent telluric features above 6860\,\AA were masked out during the fitting process. When fitting H\,$\beta$ the narrow line widths and centres were tied to that of \ion{[O}{iii]}$_{5007}$. An example fit is shown in Fig. \ref{fig:fit_example}.

\begin{table}
	\centering
	\caption{Information on the targets in this paper. The A$_{\textrm V}$ values are those obtained from \citet{Schlafly2011} assuming an R$_{\textrm V}$ of 3.1.}
	\label{tab:target_info}
	\begin{tabular}{lllll}
		\hline
		Target & RA & Dec & $z$ & A$_{\textrm V}$\\
		\hline
		J084305 & 08 43 05.55 & +55 03 51.4 & 0.8955 & 0.0824\\
		J094511 & 09 45 11.08 & +17 45 44.7 & 0.758  & 0.0718\\
		J103837 & 10 38 37.09 & +02 11 19.7 & 0.620  & 0.0877\\
		J142232 & 14 22 32.45 & +01 40 26.7 & 1.076  & 0.09\\
		\hline
	\end{tabular}
\end{table}

\section{Spectroscopic Results}
\label{sec:SpecResults}

\begin{figure*}
	\includegraphics[width=2\columnwidth]{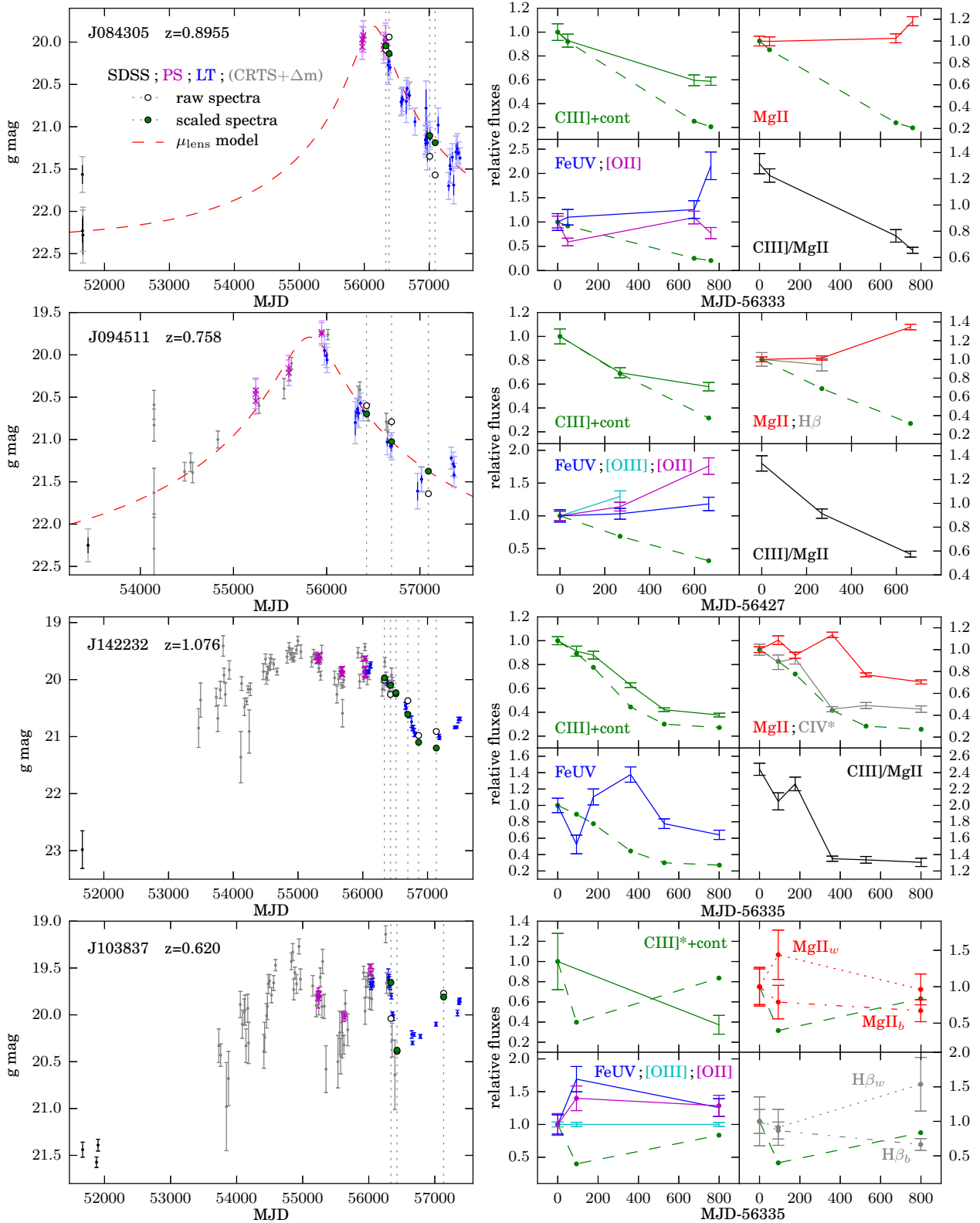}
    \caption{Left-hand panels: composite light curves for each target including spectral epochs. The open circles show the raw spectral magnitudes before rescaling as per Section \ref{sec:Pipeline} and the green circles show the rescaled values. CRTS data have offsets of +0.1/0.1/0.4/0.2 mag, respectively. The extended tails on the error bars for J084305 and J094511 reflect the errors used in the microlensing analysis. Right-hand panels: relative fluxes of the various measured spectral components. For the first three targets, the fourth panel shows the \ion{C}{iii]}/\ion{Mg}{ii} ratio. For J103837, the two Gaussian components used in the fit (broad and wide) are displayed. Asterisks denote less reliable values as noted in the text. The green dashed line reflects the continuum under \ion{C}{iii]} in each case.}
    \label{fig:LCAll}
\end{figure*}

\begin{figure*}
	\includegraphics[width=2\columnwidth]{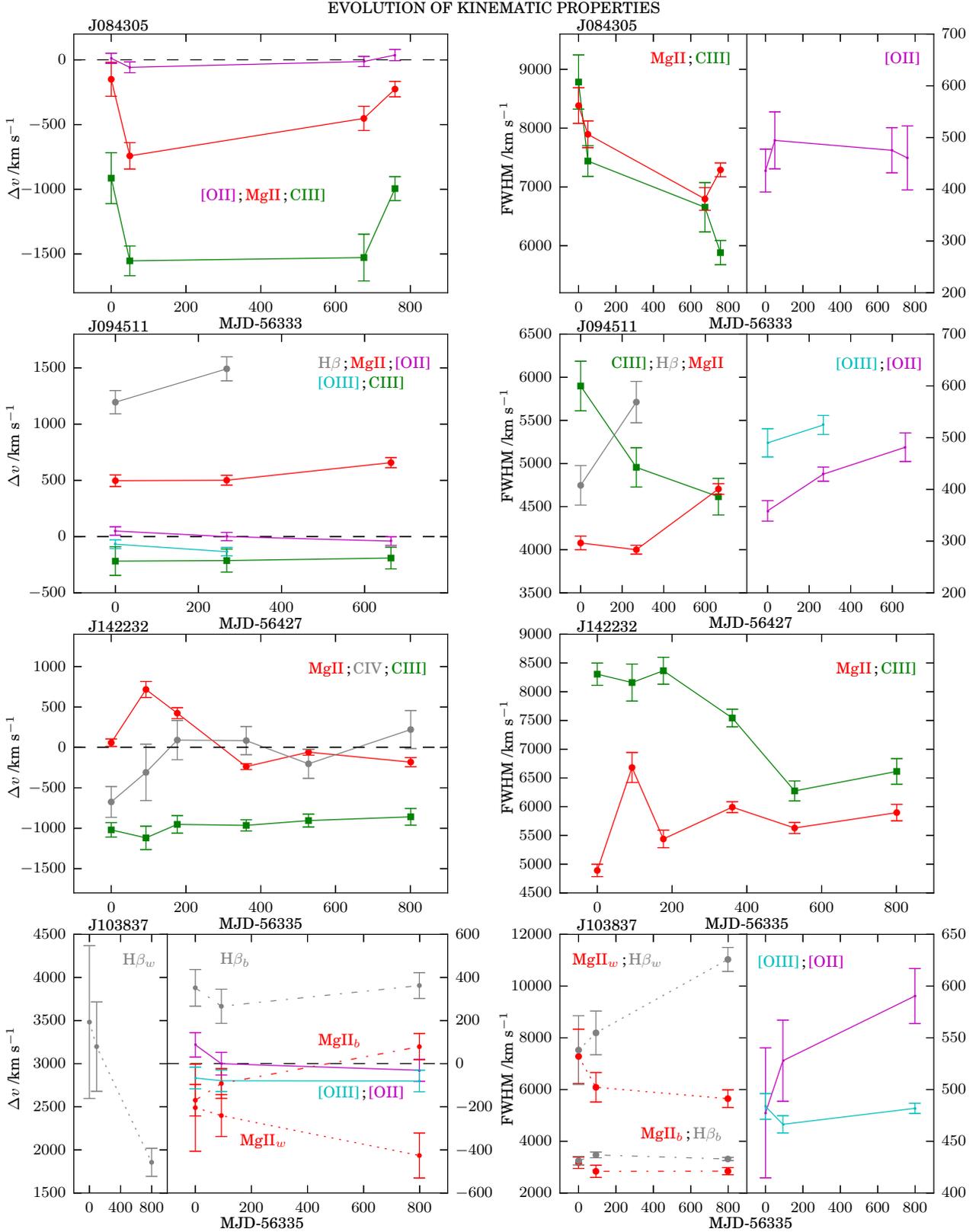}
    \caption{Left-hand panels: velocity offsets for the centre of the measured line profiles for each target assuming rest-frame wavelengths, in \AA, of the following: \ion{C}{iv}(1549); \ion{C}{iii]}(1909); \ion{Mg}{ii}(2800); \ion{[O}{ii]}(3727); H$\,\beta$/\ion{[O}{iii]}(4861/5007). The AGN systemic redshift/velocity is determined from the median of the \ion{[O}{ii]} centres. In the case of J142232, due to a lack of observed narrow lines, \ion{Mg}{ii} has been used. The error bars in this case only reflect the uncertainty in the measured line centre and omit the additional systemic velocity uncertainty. The FWHM of \ion{C}{iv} for J142232 is not shown as this value was tied to that of \ion{C}{iii]} in the fitting process. Right-hand panels: FWHM of the measured line profiles for each target.}
    \label{fig:ProfileAll}
\end{figure*}

\begin{figure*}
	\includegraphics[width=2\columnwidth]{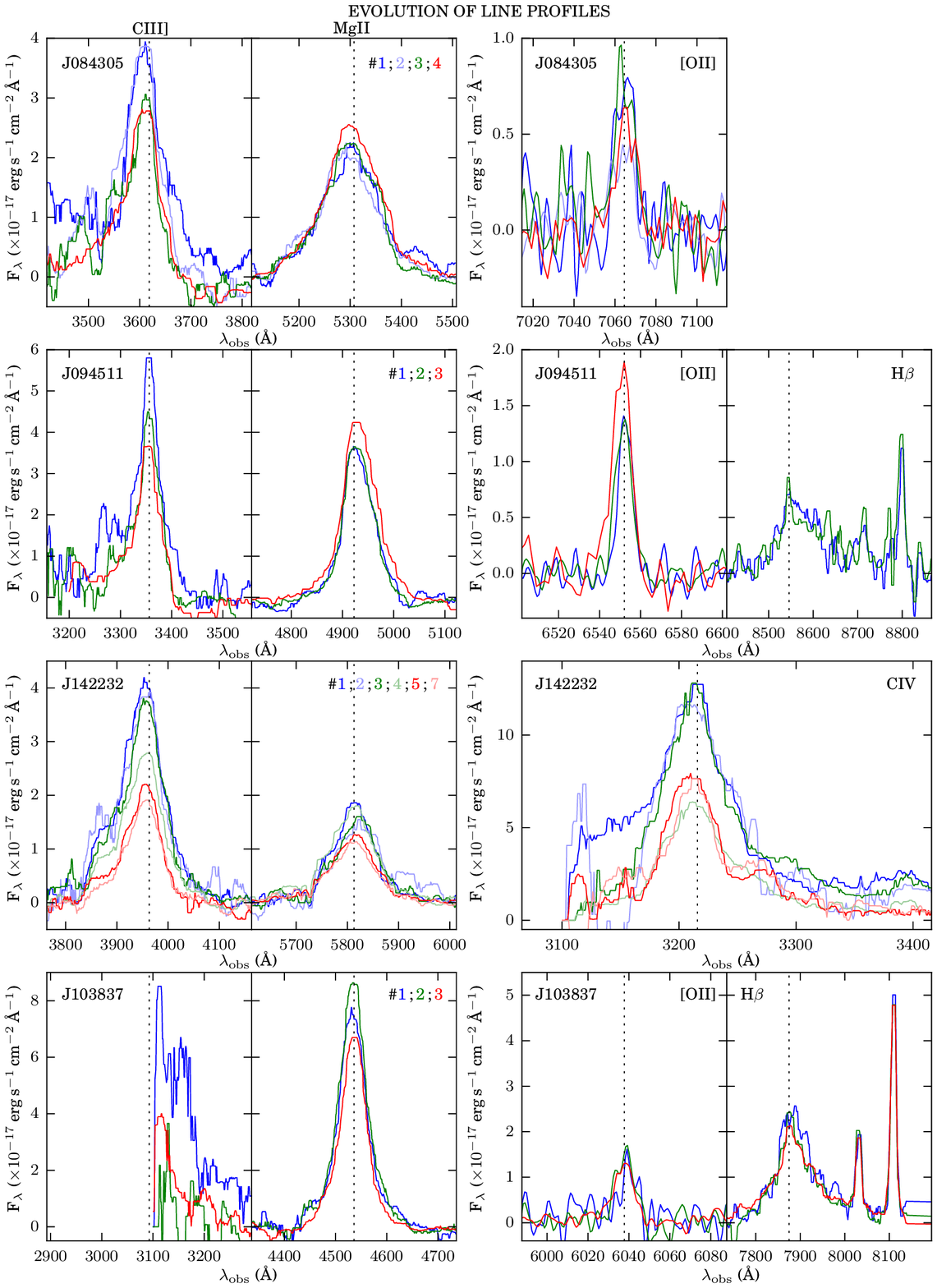}
    \caption{Left-hand panels: \ion{C}{iii]} and \ion{Mg}{ii} profiles for each object. For J103837, the profile is truncated due to the blue arm cut-off. Right-hand panels: additional emission line profiles measured in the fitting process. For clarity, all plots, barring those for \ion{[O}{ii]}, have had a median filter applied.}
    \label{fig:LinesAll}
\end{figure*}

The results from the spectroscopic analysis are presented in three key figures. Fig. \ref{fig:LCAll} shows the target light curves and line flux evolution. The line fluxes have been plotted relative to the first epoch to allow quick comparison. Fig. \ref{fig:ProfileAll} shows the evolution of the systemic velocity offsets and emission line widths. Fig. \ref{fig:LinesAll} shows the evolution of the spectral profiles after continuum and Fe template subtraction. For clarity, a median filter has been applied to the broad lines. Additional measurements can also be found in Table \ref{tab:data}. The observed spectra are shown in Figs. \ref{fig:J084305_fit}, \ref{fig:J094511_fit}, \ref{fig:J142232_fit}, \ref{fig:J103837_fit}.

Three of the four targets, J084305, J094511 and J142232, show clear evidence for a differential evolution of the continuum with respect to the line fluxes. In general, the continuum decreases by a factor $\sim4$ and, to a lesser degree, the \ion{C}{iii]} flux tracks this change. The \ion{Mg}{ii} flux either tracks the continuum change weakly or is consistent with no change at all. The photometry indicates that the targets have been evolving smoothly over this period. A more detailed summary of each object now follows and the reader should refer back to the figures mentioned in the previous paragraph.

Any evidence for an intervening galaxy, which would lend weight to the microlensing scenario, will be presented on a per target basis.

\subsection{J084305}

The light curve of this target shows a smooth decline of $\sim0.6\,\textrm{mag\,yr}^{-1}$ from a peak more than two magnitudes brighter than the SDSS observations 16 years ago. There are no CRTS data that allow us to see the pre-Pan-STARRS evolution. There are two spectral epochs near the observed maximum around 56100 MJD and a further two after a decline of at least one magnitude. The most recent photometry suggests the target is now returning to the level of the Sloan era. Of the three SDSS epochs, two are at $g\simeq22.25$ mag and the third has the higher value of $g=21.56$ mag, which may indicate a problem with the Sloan \textit{cmodel} fit(s). The microlensing model shown, to which the spectra have been rescaled, does a good job of reproducing the large amplitude changes. Further details on the model are in Section \ref{sec:MicroResultsJ084305}.

As to the spectral evolution, the continuum is seen to drop by a factor of 5 over the period spanned by the observations. The \ion{C}{iii]} flux shows a corresponding drop by a factor of $\sim2$. In contrast, the \ion{Mg}{ii} and Fe fluxes are consistent with no change or perhaps a late increase. Both the \ion{Mg}{ii} and \ion{C}{iii]} lines get narrower during the decline. In addition, both lines appear blueshifted relative to the systemic velocity, \ion{C}{iii]} more so than \ion{Mg}{ii}. There are intriguing but inconclusive signs of an evolution of the offsets, which shift bluewards by $500\,\textrm{km\,s}^{-1}$ or so and then recover. There are also apparent changes in the red wing of the \ion{Mg}{ii} profile and both wings of \ion{C}{iii]}, most notably around the expected position of the fainter \ion{Al}{iii} component. The scatter seen in the \ion{[O}{ii]} flux may be due to slit-width changes.

There are no clear spectral features suggestive of an intervening galaxy but this possibility cannot yet be ruled out. The microlensing model for this target shows that any intervening galaxy flux will be at least a factor of 2 below the AGN/host baseline flux in the $g$ band, making spectroscopic detection during an ongoing event difficult. In addition, two of the three SDSS epochs are flagged as having issues with the Petrosian radii. This may simply be due to noise or is perhaps an indicator of morphological issues.

\subsection{J094511}

Since the SDSS epoch some 11 years ago, this target has been evolving at around $\sim 0.4\,\textrm{mag\,yr}^{-1}$ and displays an approximately symmetric light curve about a peak around 56000 MJD, with an apparent dip/rise after the 57000 MJD mark. There are three spectral epochs for this object spread over two years, after the maximum, with a factor $\sim2$ decrease in luminosity over this period. As with J084305, the spectra have been scaled to the microlensing model (Section \ref{sec:MicroResults}), though the accuracy of the third epoch scaling is less certain due to the additional structure in the light curve at this point.

As with J084305, there is a marked difference between the evolution of \ion{Mg}{ii} and \ion{C}{iii]} fluxes. \ion{C}{iii]} drops to $\sim60\%$ of the initial value while \ion{Mg}{ii} is consistent with no change and possibly an increase of $\sim30\%$ if the third epoch scaling is correct. This increase is also seen for \ion{[O}{ii]}, though a larger slit width was used at the third epoch. At this epoch, the MMT spectrum did not have sufficient wavelength coverage to catch H$\,\beta$/\ion{[O}{iii]}, so it is harder to draw firm conclusions on any trends here. The H$\,\beta$ broad component is redshifted by $\sim1300\,\textrm{km\,s}^{-1}$ and \ion{Mg}{ii} shows a smaller $\sim500\,\textrm{km\,s}^{-1}$ redshift. \ion{C}{iii]} shows a modest blueshift of $\sim250\,\textrm{km\,s}^{-1}$. There is no sign of an evolution in velocity offsets as seen for J084305. The \ion{C}{iii]} and \ion{Mg}{ii} line widths show the same trend as their amplitudes, as do the narrow lines, though H$\,\beta$ only shows an increase in width. Looking at the line profiles, \ion{C}{iii]} shows a significant blue wing change whereas \ion{Mg}{ii} sees a change on the red wing. It is perhaps interesting that \ion{[O}{ii]} sees an enhancement primarily on the blue wing. No clear spectral signature of a possible intervening galaxy has been detected.

\subsection{J142232}

The light curve for this target shows there has been a rise of approximately three magnitudes since the SDSS epoch and $\lvert\Delta g\rvert\simeq1.22$ mag over the six spectral epochs as estimated from interpolation of the LT light curve. The SDSS epoch is faint and should therefore be treated with caution. The photometry shows this target to be evolving smoothly but there is a notable dip around 55500 MJD, this is after the assumed `peak' around 55000 MJD. An earlier dip around 54200 MJD is also visible though the CRTS data show considerable scatter here. After the 56500 mark, there is a rapid drop of $\sim 1.1\,\textrm{mag\,yr}^{-1}$, the time of the spectral observations, but the most recent photometry shows this levelling off. The cadence of the LT observations is not always ideal, which may affect the spectral scaling corrections (Sec. \ref{sec:Pipeline}).

For this target, outside of the rest-frame UV ($\lambda_{\rm obs}>6400$\,\AA), no other spectral features or narrow emission lines were detected. Given a lack of detected narrow lines, the systemic velocity/redshift has been determined from the median \ion{Mg}{ii} line centre. For clarity, the corresponding increase in uncertainty has been omitted from the plot in Fig. \ref{fig:ProfileAll}. Also, at this redshift the \ion{C}{iv} line is seen, albeit very near the blue cut-off. Shortward of $\sim3200$\,\AA, the flux calibration is less reliable due to the lack of calibration points for the chosen standard stars. \footnote[1]{The fitting pipeline would occasionally fail to fit \ion{C}{iv} and instead add an additional continuum component. To prevent this, the width of \ion{C}{iv} was constrained to be the same as \ion{C}{iii]}, so the \ion{C}{iv} results should be treated with caution.}

The spectroscopic results for J142232 are broadly similar to that seen in J084305/J094511. There is a factor $\sim4$ drop in the continuum and the \ion{C}{iii]}/\ion{C}{iv} fluxes appear to track this change quite closely, dropping by a factor $\sim3$. The \ion{Mg}{ii} flux also tracks the continuum to a lesser extent, dropping by a factor $\sim2$. The Fe component behaves similarly, though there is a higher degree of scatter. The \ion{C}{iii]} centre shows a blueshift of $\sim1000\,\textrm{km\,s}^{-1}$, which does not evolve, and the line gets narrower in decline. In contrast, \ion{Mg}{ii} shows a higher degree of scatter in the velocity offset and an increase in line width of $\sim1000\,\textrm{km\,s}^{-1}$ over this period. There is evidence for evolution of the line profiles, most notably on the \ion{C}{iii]} blue wing \ion{Mg}{ii}/\ion{C}{iv} red wing.

\begin{figure}
	\includegraphics[width=\columnwidth]{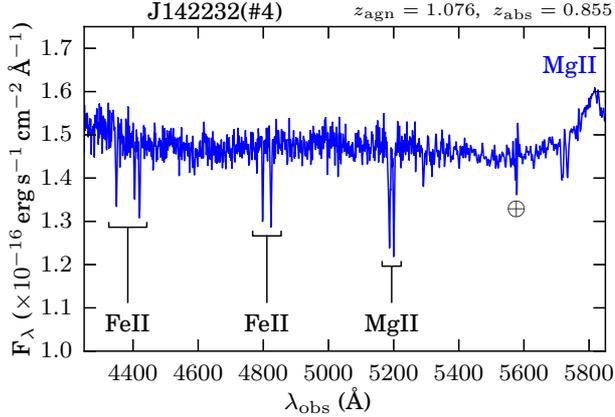}
    \caption{Fourth spectral epoch for J142232 highlighting the narrow absorption features seen. These are consistent with \ion{Mg}{ii}/FeII absorption at a lower redshift or possibly an outflow intrinsic to the AGN. A similar \ion{Mg}{ii}/FeII feature is seen nearer the blue wing of the \ion{Mg}{ii} broad line though in this case only the \ion{Mg}{ii} doublet is readily visible.}
    \label{fig:J142232_abs}
\end{figure}

In addition, there is also a set of strong narrow absorption features present, possibly evidence for an intervening galaxy. These features are consistent with \ion{Mg}{ii}/Fe absorption at $z=0.855$ and are highlighted in Fig. \ref{fig:J142232_abs}. The AGN is at $z=1.076$. Alternatively, this feature may be the result of an outflow intrinsic to the AGN. However, this $\Delta v$ is close to 0.1$c$, much greater than the value of 0.01$c$ typically used to distinguish between associated and intervening systems \citep{York2006} making an outflow unlikely in this case. There is also a second similar but fainter feature, seen nearer the blue wing of the \ion{Mg}{ii} broad line. In addition, the SDSS epoch for this target is flagged as a possible blend, i.e. more than one peak was detected in a given filter. Given that the target was faint at this epoch, this may not be reliable but lends further weight to the possible presence of an intervening galaxy.

\subsection{J103837}

This object exhibits the most complex light curve and the spectra have been scaled relative to the \ion{[O}{iii]}$_{5007}$ flux (Section \ref{sec:Pipeline}). There is evidence of at least three peaks with occasional rapid changes in brightness, particularly around 56300 MJD, which shows a drop of $\sim0.4$\,mag over 2.5 months. Here, the cadence of the LT photometry is less than ideal. It is possible the drop was larger than observed. After the drop, it appears to rise more gradually at $\sim0.2\,\textrm{mag\,yr}^{-1}$ though there is a spike in the data ($g=19.79$ mag) at the third spectral epoch at 57137 MJD. There is a peak-to-peak change from the Sloan era of at least two magnitudes. The increase in complexity, coupled with the lack of LT data at the second spectral epoch causes a problem when attempting to correct the spectral data for transparency effects (Section \ref{sec:Pipeline}). For this object, the \ion{[O}{iii]} flux was used to rescale the data as the interpolation using the LT data was poor. This may introduce additional errors if the line flux varies either intrinsically over this time-scale or due to aperture/seeing effects.

Given that in this case the second spectral epoch represents the minimum, it is harder to draw conclusions regarding any spectroscopic trends. The SNR of the third epoch was sufficient to allow a more complex fit for the H$\,\beta$/\ion{Mg}{ii} lines. In this case, two Gaussian profiles were used simultaneously. The \ion{C}{iii]} results should be treated with caution as the bulk of the line was beyond the blue cutoff requiring that the line centre and width be fixed relative to that of \ion{Mg}{ii}. It was not reliably detected at the second epoch. Both of the wider Gaussian components for H$\,\beta$/\ion{Mg}{ii} appear to undergo a blueshift over the period of observations and H$\,\beta$ shows evidence of an increase in FWHM also.

\begin{figure}
	\includegraphics[width=\columnwidth]{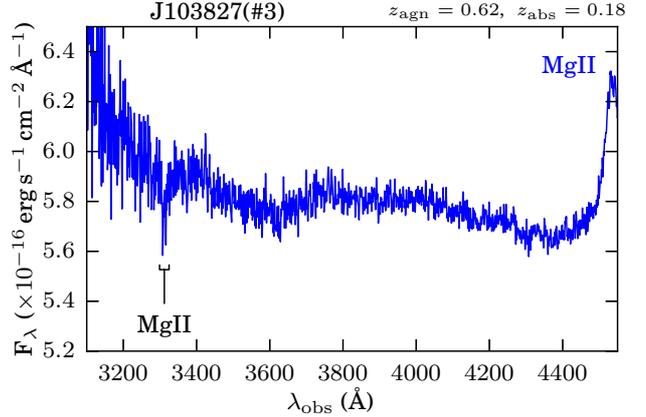}
    \caption{Third spectral epoch for J103837 highlighting the narrow absorption feature. It is consistent with an \ion{Mg}{ii} doublet at a lower redshift than the AGN.}
    \label{fig:J103837_abs}
\end{figure}

For this target, an absorption doublet was seen that may indicate the presence of an intervening galaxy. It is consistent with \ion{Mg}{ii} at a low redshift and is highlighted in Fig. \ref{fig:J103837_abs}. This feature has only been reliably detected in the third epoch, likely due to the higher SNR achieved in the blue. Here, the suspected absorber lies at $z=0.18$ and is far less likely to be an outflow intrinsic to the AGN ($z=0.62$) as might be the case for the absorption seen in J142232. If in outflow, the velocity would be in excess of 0.25$c$ (i.e. $>>0.01c$). This is seen in some high-ionization species but is perhaps implausibly high for these low-ionization lines.

\section{Microlensing Models: Techniques}
\label{sec:MicroModels}

This section will summarize the methods used for exploring the suggestion that these AGN transients are microlensing events. The results will be detailed in Section \ref{sec:MicroResults}.

We concentrate on modelling events due to single isolated lensing stars, rather than the complex magnification maps due to multiple stars in the line of sight normally considered in quasar microlensing studies (e.g. \citealt{Wambsganss1992,Lewis1995,Vernardos2013}). This is justified because we are dealing with random sightlines rather than objects pre-selected as multiply imaged quasars. For example a sight line through the Milky Way at the solar radius, with stellar density $\sim 0.1$ pc$^{-3}$ and scaleheight 300 kpc, would produce $\kappa\sim 0.2$ if placed at $z=0.2$ with the quasar at $z=1.0$. In fact, as explained in L16, we will usually be looking through a galaxy well below $L_*$.

On the other hand, the effect of shear caused by the overall potential of the lensing galaxy containing the microlensing star, will often be significant, so that in general we will be dealing with `Chang--Refsdal' (CR) lenses \citep{Chang1984}. For two of our targets, with double peaks (J103837 and J142232), the shear is clearly important. The other two (J094511 and J084305) are smooth and single peaked, so we start with the simplest model ignoring shear and using a simple point lens. In Sections \ref{sec:Extended_sources} and \ref{sec:CR}, we examine the effect of varying that assumption.

\subsection{Point-source point-lens model}

It is useful to first start with the simplest of cases, that of a point-mass lens and point-source with no external shear. This is likely an oversimplification but will nevertheless prove useful, particularly with regard to J084305 and J094511, both of which have light curves that are single-peaked and smoothly evolving (Fig. \ref{fig:LCAll}). In this model the magnification is given by:

\begin{equation}
	\mu=\frac{y^2+2}{y\sqrt{y^2+4}},\quad
	y:=\beta/\theta_\textrm{E}.
\end{equation}

Here, $y$ is the normalized source position in units of the Einstein radius of the lens, which is given by:

\begin{equation}
	\theta_\textrm{E}=\left(\frac{4GM}{c^2}\frac{D_\textrm{ds}}{D_\textrm{d}D_\textrm{s}}\right)^{1/2}.
	\label{eq:theta_E}
\end{equation}

$D_\textrm{d}$, $D_\textrm{s}$ and $D_\textrm{ds}$ are the angular diameter distances for the lens, source and between the lens and source, respectively. To compute the light curve $F(t)=\mu(t)F_\textrm{s}$ requires the formula for the trajectory of the source, as in \citet{Wambsganss2006}:

\begin{equation}
	y(t)=\sqrt{y_0^2 + \left(\frac{t-t_0}{t_\textrm{E}}\right)^2},\quad
	t_\textrm{E}:=\frac{D_\textrm{d}\theta_\textrm{E}}{v_\perp},
	\label{eq:y(t)}
\end{equation}

where $y_0$ is the impact parameter at $t_0$ and $v_\perp$ is the transverse velocity of the lens relative to the observer/source line of sight. The Einstein time-scale, $t_\textrm{E}$ defines a characteristic time-scale for the lensing event. If one includes an additional background flux contribution from the host/lens galaxy, this model has seven free parameters. These are listed in Table \ref{tab:params}.

\begin{table}
	\centering
	\caption{Free parameters in the simple microlensing model.}
	\label{tab:params}
	\begin{tabular}{cl}
		\hline
		Parameter & Description \\
		\hline
		$M_\textrm{l}$ & Lens mass \\
		$v_\perp$ & Transverse velocity \\
		$t_0$ & Mid-point epoch \\
		$z_\textrm{d}$ & Lens redshift \\
		$y_0$ & Impact parameter \\
		$F_\textrm{s}$ & Source flux (pre-lensing) \\
		$F_\textrm{b}$ & Background flux (unlensed) \\
		\hline
	\end{tabular}
\end{table}

In order to explore this model in more detail, \textsc{emcee}, a Bayesian model fitting package that uses a Markov chain Monte Carlo (MCMC) method to explore parameter space, was used \citep{Foreman-Mackey2013}. Gaussian errors are assumed and the parameters constrained to remain physical. In order to deal with the mass/velocity degeneracy in the model, a lognormal prior on the expected lens mass was used. This takes the form of a Chabrier IMF \citep{Chabrier2003} with an additional weighting factor $M$ to account for the increase in lensing cross-section with mass. The lower limit for the lens mass was set at $10^{-3}$M$_\odot$ and the transverse velocity of the lens was left free. Over the course of a lensing event, an AGN will also have some level of, as yet unknown, intrinsic variability. A simple approach that attempts to allow for this was to increase the errors on the photometry by an additional 10\% of the flux level at each epoch, a fairly conservative estimate for typical AGN variability. The CRTS data were not used in this analysis due to concerns over colour effects resulting from the clear filter used.

Parameter constraints were obtained using the marginalized posterior probability distributions. For each, the peak value was determined by fitting a polynomial through the maximum of the distribution using a coarse sampling (logarithmic bins for $M_\textrm{l}$ and $v_{\perp}$). A finer sampling of the distribution was used to determine the narrowest allowable range, which encompassed $>68$ per cent of the data about this peak. In addition, values for $r_\textrm{E}$, the Einstein radius in the source plane, were calculated from the MCMC trace output. This distribution (logarithmic bins) was then used to produce constraints in the same manner as for the other parameters. The results from our MCMC analysis will be displayed in Section \ref{sec:MCMCresults}.

\subsection{Extended sources}
\label{sec:Extended_sources}

The assumption of the AGN as a point-source will not always be valid, so it is necessary to consider extended source models. Indeed, this fact can potentially be exploited to yield additional information regarding the accretion disc and Broad Line Region (BLR) structure. In order to explore extended source models in more detail, an inverse-ray-shooting code technique \citep{Jimenez-Vicente2016} was used to construct magnification maps, initially for a point-lens model. The maps that have been constructed are normalized in units of the Einstein radius, have a pixel scale of 1/800\,$\theta_\textrm{E}$ and side length of 4\,$\theta_\textrm{E}$. A simplifying assumption is that the accretion disc and BLR sources are Gaussian. This should be reasonable for low inclinations ($\phi<45^\circ$) and disc-like morphologies, when projected along the line of sight \citep{Mortonson2005,Sluse2011}. For a given set of lens/source parameters, the light curve for the event is obtained by varying the source position under the magnification map as a function of time and integrating to find the total amplification at each epoch.

While we have yet to perform a detailed MCMC analysis of extended sources under a point-mass lens, as we have done for the point-source model, we are able to use this extended model to determine whether or not the point-source approximation remains valid when estimates for the true disc size are used. Our starting point for this testing is first to derive estimates for the expected accretion-disc size. For this, we make use of the simplified thin-disc model from \citet{Morgan2010}. This provides a simple formula for the effective disc size:

\begin{equation}
	{\rm log_{10}}\left(\frac{R_{2500}}{{\rm cm}}\right)=15.184+\frac{2}{3}{\rm log_{10}}\left(\frac{M_{\rm BH}}{10^9{\rm M_\odot}}\right)+\frac{1}{3}{\rm log_{10}}\left(\frac{L_{\rm bol}}{\eta L_{\rm E}}\right),
	\label{eq:thindisc}
\end{equation}

where $L_{\rm bol}$ is the bolometric luminosity, $\eta$ the accretion efficiency and $L_{\rm E}$ the Eddington luminosity. However, from the study of differential microlensing of components of strongly lensed quasars, \citet{Morgan2010} also found evidence that the accretion disc is larger than this relationship implies, typically by a factor of a few. They derive the following empirical relationship:

\begin{equation}
	{\rm log_{10}}\left(\frac{R_{2500}}{{\rm cm}}\right)=15.78+0.8\,{\rm log_{10}}\left(\frac{M_{\rm BH}}{10^9{\rm M_\odot}}\right).
	\label{eq:Morgan}
\end{equation}

Both of these relationships require an estimate for the mass of the black hole. For this, we make use of the \citet{McLure2004} single epoch relation:

\begin{equation}
	\frac{M_{\rm BH}}{{\rm M_\odot}}=3.2\left(\frac{\lambdaup L_{3000}}{10^{37}{\rm W}}\right)^{0.62}\left[\frac{{\rm FWHM_{\ion{Mg}{ii}}}}{{\rm km\,s^{-1}}}\right]^2.
	\label{eq:McLure}
\end{equation}

This gives the black hole mass as a function of \ion{Mg}{ii} line width and luminosity at 3000\,\AA. This relationship depends on accurate estimates of the monochromatic continuum luminosity and \ion{Mg}{ii} width. Both may be overestimated if there is an ongoing microlensing event. In the absence of a confirmed AGN baseline luminosity, we make use of the faintest(latest) spectral epoch for these calculations. The black hole mass estimates for each epoch are listed in Table \ref{tab:data}. For equation (\ref{eq:thindisc}), the bolometric luminosity was calculated using the monochromatic luminosity ($\lambdaup L_{2500}$), from the same epoch as the black hole mass estimate, along with a bolometric correction factor of 5.6 \citep{Elvis1994}. The accretion efficiency was assumed to be 0.1. Finally, the Gaussian sources for testing with the magnification maps are generated with a $2\sigma$ radius equivalent to the accretion-disc size estimates obtained above. 

An important test of our extended source model is to see if it can reproduce the analytic light curve from the point-source model in the limit when the source size is very small. This is achieved by setting the size of the source to a single pixel and testing a wide range of impact parameters. The results are in excellent agreement, even for very small impact parameters, and implies that pixel/resolution effects should be minimal as long as source sizes larger than the pixel scale are used.

Sources larger than the accretion-disc estimates, which are more appropriate when considering BLR emission, are also considered. The results of the extended source analysis are displayed in Section \ref{sec:ExtendedResults}.

\subsection{Towards more realistic lens models}
\label{sec:CR}

For the models we have discussed so far, the assumption of a point-mass lens is reasonable as it is assumed the microlensing is caused by a single star. However, this neglects the presence of an external shear due to the lens-host galaxy or additional stars along the line of sight, which can have significant consequences. This background perturbation breaks the circular symmetry of the point-mass lens and results in the degenerate point in the source plane unfolding into a caustic pattern of varying complexity. The corresponding light curve may exhibit double peaks or more complex structure as a result, particularly if the size of the source is small enough relative to the caustic network.

In multiply imaged quasars, it is possible to use the positions of the quasar images and lens galaxy to estimate the convergence and shear parameters ($\kappa$,$\gamma$) for each image. This also encapsulates the expected macromagnification before any microlensing perturbations are taken into account. However, in our case, we do not yet have an unambiguous detection of the position/redshift of any lens galaxy or multiple images, which may be going unresolved in the photometry. With this in mind, the point-lens models mentioned previously have the implicit working assumption that the convergence and shear, and hence macromagnification, are small enough to be neglected. This is likely to be an oversimplification but should be sufficient to demonstrate the feasibility of the microlensing scenario for the single-peaked objects. This will be discussed in sec. \ref{sec:LensDiscuss}.

For the multiply peaked objects, a more complex lensing model is required. A thorough treatment of the many possible lens configurations for these objects is beyond the scope of this paper but we now include a simplified model to allow a qualitative exploration. The model, which describes the effect of microlensing plus an external shear, is that of the CR lens \citep{Chang1984}. This is a useful starting point as it treats the lens-host galaxy as an additional point mass nearby. Here, the shear parameter is defined as the ratio of the Einstein radius of the galaxy to the distance of the stellar lens from the galaxy centre, all squared. If this ratio is greater than unity, there can be strongly demagnified regions in the lens map. This adds a further layer of complexity when considering the light curves. For very small shear values, the magnification map is essentially the same as for an isolated point lens.

\begin{figure}
	\includegraphics[width=\columnwidth]{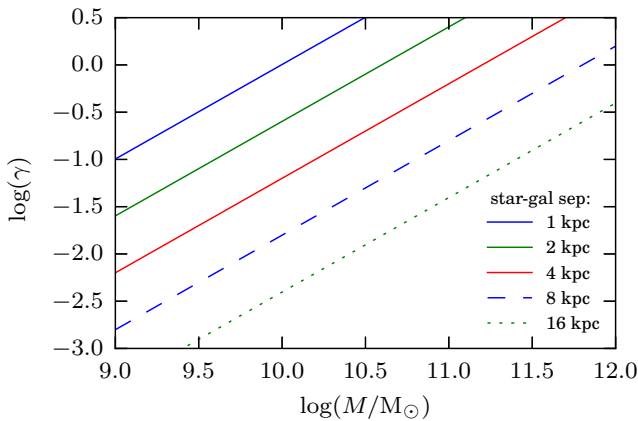}
    \caption{Plot of the Chang-Refsdal shear parameter, $\gamma$, against galaxy mass for selected star-galaxy distances. Here, the lens redshift is assumed to be 0.2 and the source redshift 1.}
    \label{fig:CRshear}
\end{figure}

Fig. \ref{fig:CRshear} shows how the shear varies in relation to the lens galaxy mass for selected star--galaxy separations. This assumes a lens redshift of 0.2 and source redshift of 1.0. L16 estimates that an AGN will have a foreground galaxy present in perhaps 0.2\% of cases but only a small fraction of these will have an ongoing high-amplitude microlensing event at any one time. We appear to be in a regime where some events will be well approximated by a point lens and others will require a more detailed treatment of the perturbing lens galaxy.

The use of a CR magnification map allows a qualitative explanation for the double-peaked light curve seen for J142232 and the result is displayed in Section \ref{sec:J142232_Extended}. An analysis of more complex lensing models for these objects is deferred to a later paper.

\section{Microlensing Models: Results}
\label{sec:MicroResults}

In this section, the results from the analysis of the various microlensing models are presented. First is the point-source point-lens model MCMC analysis. This performs well for the two single-peaked targets, J084305 and J094511. Secondly, extended source models are applied to the same targets. We find it is not possible to differentiate between extended accretion-disc models and the point-source approximation, suggesting the disc may be only marginally resolved in these cases. Also, while not a full exploration of parameter space, we are able to place some constraints on the size of the emitting region for \ion{C}{iii]}, particularly for J084305. Finally, an initial exploration into the more complex case of microlensing with an external shear is made for J142232 and J103837.

\subsection{Point-source point-lens model}
\label{sec:MCMCresults}

\subsubsection{J084305}
\label{sec:MicroResultsJ084305}

The microlensing parameter estimates for J084305 are shown in Table \ref{tab:J084305_params} and the  marginalized parameter model fit to the data is shown in Fig. \ref{fig:LCAll}. The full MCMC results are displayed in Fig. \ref{fig:J084305_corner}.  Here, one can see that there is a strong correlation of lens mass with transverse velocity, as expected, and also with impact parameter versus the pre-lensed AGN (or `source') flux. An anticorrelation is seen for the background flux with respect to both impact parameter and source flux. The model light curve produces a fit to the data with a reduced chi-squared value of 0.69. This assumes a negligible background component and has been calculated using the same photometric errors as that for the MCMC analysis. Given the good fit to the data, this model was used to scale the spectra. This specific model has an Einstein radius of 12.1 light-days, Einstein time-scale of 7.5 yr and a peak amplification of 10.3.

Though poorly constrained, the values obtained for the lens mass (0.37 M$_\odot$), transverse velocity ($830\,\textrm{km\,s}^{-1}$) and redshift (0.34 as compared to 0.895 for the AGN) are all physically reasonable. The estimate for the background flux from the host/lens galaxies is consistent with an upper limit ($g\leq23.6$ mag) and the AGN would have an unlensed magnitude of $g\simeq22.3$ mag.

\begin{table}
	\centering
	\caption{Parameter estimates obtained from the MCMC analysis, including the Einstein radius in the source plane, for J084305. The model using these marginalized parameters, assuming a negligible background flux, produces a fit to the data with a reduced chi-squared value of 0.69 and has an Einstein radius of 12.1 light-days. Parameter definitions are as per Table \ref{tab:params}.}
	\label{tab:J084305_params}
	\begin{tabular}{clc}
    	\hline
		J084305 & & $z_{\rm agn}=0.8955$ \\
		\hline
		parameter & value & unit \\
		\hline
		$z_d$     & $0.34\,^{+0.27}_{-0.21}$ & \\
		$M_l$     & $0.37\,^{+1.19}_{-0.29}$ & M$_\odot$ \\
		$v_\perp$ & $830\,^{+1080}_{-510}$ & km\,s$^{-1}$ \\
		$y_0$     & $0.097\,^{+0.026}_{-0.031}$ & $\theta_E$\\
		$t_0$     & $56137\,^{+23}_{-23}$ & MJD \\
		$F_s$     & $5.5\,^{+1.5}_{-1.6}$ & $\times10^{-18}$erg\,s$^{-1}$cm$^{-2}$\AA$^{-1}$ \\
		$F_b$     & $<1.8$ & $\times10^{-18}$erg\,s$^{-1}$cm$^{-2}$\AA$^{-1}$\\
		$r_E$     & $10.5\,^{+17.0}_{-6.3}$ & light-days \\
		& $r_{E,{\rm mp}}=12.1$ l.d. & $\chi^{2}_{\nu}=0.69$ \\
        \hline
	\end{tabular}
\end{table}

\subsubsection{J094511}

The microlensing parameter estimates for J094511 are shown in Table \ref{tab:J094511_params} and the marginalized parameter model fit to the data is shown in Fig. \ref{fig:LCAll}. The model light curve produces a fit to the data with a reduced chi-squared value of 1.14. This specific model has an Einstein radius of 12.8 light-days, Einstein time-scale of 10.8 yr and a peak amplification of 13.5. The CRTS data was not used in the analysis but is in broad agreement with the model. The accuracy of the model may be adversely affected due to the extra structure seen in the light curve at later epochs (MJD$>$57000).

Similarly to J084305, the same correlations are seen and the parameter estimates are physically reasonable. In this case, the transverse velocity and impact parameter are lower. Using these parameter estimates, the Einstein time-scale is 10.8 yr, Einstein radius 12.8 light-days and the peak amplification factor is 13.5. The estimate for the unlensed flux level from the host/lens galaxies is consistent with an upper limit ($g\leq23.6$ mag) and the AGN would have an unlensed magnitude of $g\simeq22.6$ mag.

\begin{table}
	\centering
	\caption{Parameter estimates obtained from the MCMC analysis, including the Einstein radius in the source plane, for J094511. The model using these marginalised parameters, assuming a negligible background flux, produces a fit to the data with a reduced chi-squared value of 1.14 and has an Einstein radius of 12.8 light-days. Parameter definitions are as per Table \ref{tab:params}.}
	\label{tab:J094511_params}
	\begin{tabular}{clc}
    	\hline
		J094511 & & $z_{\rm agn}=0.758$ \\
		\hline
		Parameter & Value & Unit \\
		\hline
		$z_\textrm{d}$     & $0.29\,^{+0.24}_{-0.18}$ & \\
		$M_\textrm{l}$     & $0.40\,^{+1.47}_{-0.32}$ & M$_\odot$ \\
		$v_\perp$ & $580\,^{+950}_{-360}$ & km\,s$^{-1}$ \\
		$y_0$     & $0.074\,^{+0.033}_{-0.029}$ & $\theta_\textrm{E}$\\
		$t_0$     & $55816\,^{+26}_{-26}$ & MJD \\
		$F_\textrm{s}$     & $4.3\,^{+1.8}_{-1.8}$ & $\times10^{-18}$erg\,s$^{-1}$cm$^{-2}$\,\AA$^{-1}$ \\
		$F_\textrm{b}$     & $<1.8$ & $\times10^{-18}$erg\,s$^{-1}$cm$^{-2}$\,\AA$^{-1}$\\
		$r_\textrm{E}$     & $11.9\,^{+18.9}_{-7.2}$ & Light-days \\
		& $r_{{\rm E, mp}}=12.8$ l.d. & $\chi^{2}_{\nu}=1.14$ \\
        \hline
	\end{tabular}
\end{table}

\subsection{Extended source models}
\label{sec:ExtendedResults}

The preceding section provides good evidence that the point-source, point-lens microlensing model is a reasonable approximation for the continuum changes seen in J084305 and J094511. We now explore these lensing models in the regime where the point-source approximation no longer applies. This is particularly important with regards to the size of the BLR. The spectroscopic analysis shows there is evidence for the differential evolution of the BLR fluxes, namely \ion{Mg}{ii} and \ion{C}{iii]}, with respect to the continuum. If these lines are partially resolved by the lens, then it allows constraints to be placed on the size of the emitting region. Indeed, in the case of J084305 in particular, there is evidence to suggest that the \ion{C}{iii]} region is compact enough to undergo significant changes in amplification as the lensing event unfolds.

\subsubsection{J084305}
\label{sec:J084305_Extended}

\begin{figure}
	\includegraphics[width=\columnwidth]{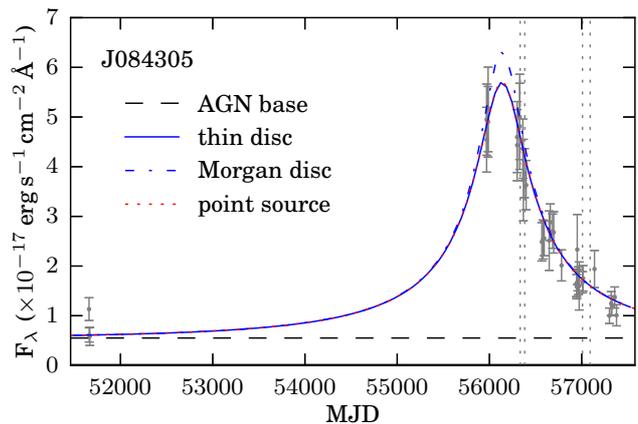}
    \caption{Extended source model light curves for J084305. The grey data are the original photometry points and the vertical lines are the spectral epochs. The source sizes are listed in Table \ref{tab:J084305ADsizes}. The dashed line represents the AGN base flux.}
    \label{fig:J084305AD}
\end{figure}

In order to test the validity of the point-source model, we compare extended accretion-disc models using empirical estimates for the disc size and contrast with the point-source approximation. The lens model parameters and accretion-disc size estimates are listed in Tables \ref{tab:J084305_params} and \ref{tab:J084305ADsizes} respectively. The results are shown in Fig. \ref{fig:J084305AD}. It is clear from the figure that there is little separation between the models except perhaps during the midpoint of the event. Here, the increased amplification seen at the peak for the larger Morgan disc is likely a result of the source being of a suitable size to have significant flux overlapping the regions of highest magnification for the given impact parameter. The fit to the light curve for the thin-disc model has $\chi^2_{\nu}=0.70$ and for the Morgan disc $\chi^2_{\nu}=0.88$. For lensing models with a smaller projected Einstein radius (varying $z_\textrm{d}$, $M_\textrm{l}$, $v_\perp$ within the derived parameter constraints), the distinction between the models becomes more apparent, favouring the smaller thin-disc and point-source models. Conversely, for larger Einstein radii, both extended models converge to the point-source solution.

The fact that both the point-source and extended accretion-disc models perform well suggests that, within the constraints imposed by the light curve, the point-source approximation is reasonable, at least in the $g$ band. However, the possibility that the disc is being resolved, if only marginally, still exists. This does not rule out chromatic effects due to microlensing but makes it harder to draw any conclusions as to the true thermal profile of the disc.

\begin{table}
	\centering
	\caption{Empirical estimates for the accretion-disc sizes for J084305, calculated as per Section \ref{sec:Extended_sources}.}
	\label{tab:J084305ADsizes}
	\begin{tabular}{lll}
		Source & Size        &                   \\
		       & (light-days) & ($\theta_{\rm E}$) \\
		\hline
		Thin disc     & 0.19 & 0.016 \\
		Morgan disc   & 0.91 & 0.075 \\
	\end{tabular}
\end{table}

\begin{table}
	\centering
	\caption{Estimated \ion{C}{iii]} and \ion{Mg}{ii} sizes for J084305. They have been determined from the spectroscopic data as per Section \ref{sec:J084305_Extended}. The errors here reflect the fitting process only.}
	\label{tab:J084305BLRsizes}
	\begin{tabular}{lll}
		Source & Size        &                   \\
		       & (light-days) & ($\theta_{\rm E}$) \\
		\hline
		\ion{C}{iii]}		  & $4.8\pm0.2$  & $0.40\pm0.02$ \\
		\ion{Mg}{ii}          & $>15$        & $>1.24$       \\
	\end{tabular}
\end{table}

\begin{figure}
	\includegraphics[width=\columnwidth]{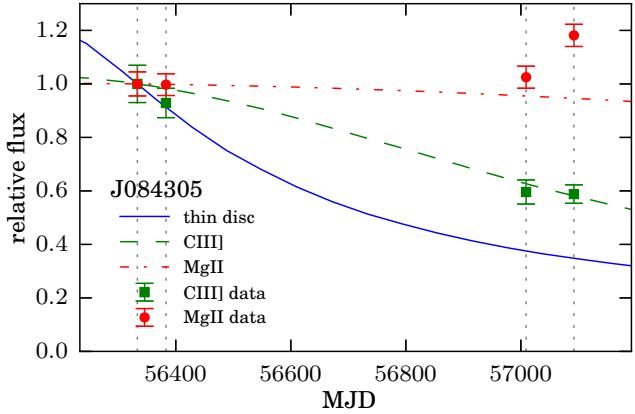}
    \caption{Evolution of extended model fluxes relative to the first spectral epoch. Overplotted are the spectroscopic measurements of \ion{C}{iii]} and \ion{Mg}{ii}. The source sizes are listed in Table \ref{tab:J084305BLRsizes}.}
    \label{fig:J084305BLR}
\end{figure}

There are indications that the BLR is being partially resolved by the lens. Fig. \ref{fig:J084305BLR} shows the flux evolution for selected source models, relative to the first spectroscopic epoch, against the observed data for \ion{C}{iii]} and \ion{Mg}{ii}. Again, note the differential evolution of the lines with respect to the continuum. If one assumes that this is a lensing effect, the data for \ion{C}{iii]} is sufficient to allow a determination of the size of the emitting region using a $\chi^2$ distribution, assuming Gaussian errors. The values are noted in Table \ref{tab:J084305BLRsizes}. Given the uncertainty in the Einstein radius of the lens, the relative size of 0.4 $\theta_{\rm E}$ should prove more robust than the absolute size. For this particular lens model, the absolute size of the \ion{C}{iii]} emitting region is $4.8\pm0.2$ light-days. The true value is likely to be in the range $\sim2\textup{--}11$ light-days when the uncertainties in the lens model parameters are taken into account. In the case of \ion{Mg}{ii}, the data is consistent with no change or perhaps a late increase in flux. A source size greater than approximately 15 light-days (or > 1.2 $\theta_{\rm E}$) would see little evolution in flux over this period. This provides a lower limit to the size of the \ion{Mg}{ii} emitting region. It does not explain the jump in flux seen for the fourth epoch. However, this could be due to a scaling problem. As will be discussed below, the estimate for the size of the \ion{C}{iii]} emitting region is relatively compact when compared with size predictions for the \ion{C}{iv} and H$\,\beta$ regions.

\subsubsection{J094511}
\label{sec:J094511_Extended}

J094511 shows similarities to that of J084305. The empirical accretion-disc size estimates are listed in Table \ref{tab:J094511sizes}. Using the lensing model parameters in Table \ref{tab:J094511_params}, these provide light curves that are nearly indistinguishable from the point-source case. Again, this suggests that the point-source approximation is reasonable but does not rule out the possibility that the disc is being resolved. For this target, the uncertainty on the scaling of the third epoch means that it is not yet possible to place reasonable constraints on BLR sizes. What can be said for this object is that if the observed changes are due to microlensing, then the size of the \ion{C}{iii]} region again has to be small, perhaps of the order of $\sim$1 light-day as this line appears to track the continuum closely.

\begin{table}
	\centering
	\caption{Empirical estimates for the accretion-disc sizes for J094511, calculated as per Section \ref{sec:Extended_sources}.}
	\label{tab:J094511sizes}
	\begin{tabular}{lll}
		Source & Size        &                   \\
		       & (light-days) & ($\theta_{\rm E}$) \\
		\hline
		Thin disc     & 0.097 & 0.0076 \\
		Morgan disc   & 0.31 & 0.024 \\
	\end{tabular}
\end{table}

\subsection{Lensing plus external shear}

\subsubsection{J142232}
\label{sec:J142232_Extended}

A detailed analysis of more complex lensing morphologies is beyond the scope of this paper, but an exploratory analysis is presented here in order to address the question of whether or not the double- or indeed multiply peaked light curves seen in this sample can also be due to microlensing.

\begin{figure}
	\includegraphics[width=\columnwidth]{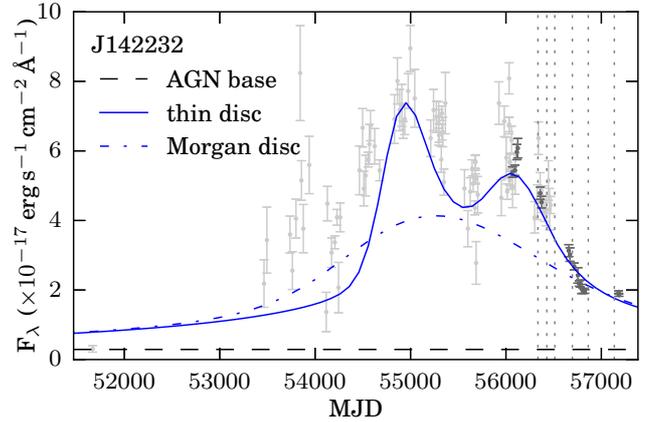}
    \caption{Chang-Refsdal lens model light curves for J142232. The grey data are the original photometry points, the darker points are the LT epochs. The source sizes are listed in Table \ref{tab:J142232sizes}. The dashed line represents the AGN base flux.}
    \label{fig:J142232AD}
\end{figure}

\begin{figure}
	\includegraphics[width=\columnwidth]{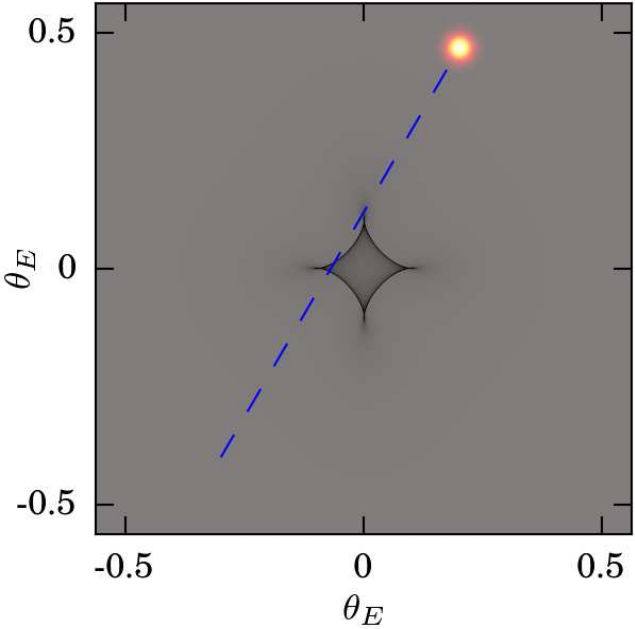}
    \caption{Magnification map and source/track to produce the thin-disc light curve shown in Fig. \ref{fig:J142232AD}.}
    \label{fig:J142232map}
\end{figure}

J142232 exhibits a clear double-peak in the light curve, which requires the use of a more complicated magnification map in order to try and reproduce this. For this test, a CR lens (Sec. \ref{sec:CR}) with the external shear parameter set to 0.05 was used. With reference to Fig. \ref{fig:CRshear}, this value is reasonable for smaller galaxies with a star--galaxy separation $\sim$4 kpc. It is enough to unfold the central degenerate point into a diamond-like caustic region. For this target, there is also a possible lens redshift due to the presence of absorption features in the spectrum. Using this and an assumed lens mass of 0.5 ${\rm M_\odot}$ gives this particular lens model an Einstein radius at source of $\sim$6.24 light-days. As for J084305, two empirical accretion-disc size estimates were used and these sizes noted in Table \ref{tab:J142232sizes}. Next, the source track and intrinsic flux were varied to provide the qualitative light curves seen in Fig. \ref{fig:J142232AD}. The magnification map and source/track used in this model are shown in Fig. \ref{fig:J142232map}.

\begin{table}
	\centering
	\caption{Empirical estimates for the accretion-disc sizes for J142232, calculated as per Section \ref{sec:Extended_sources}.}
	\label{tab:J142232sizes}
	\begin{tabular}{lll}
		Source & Size        &                   \\
		       & (light-days) & ($\theta_{\rm E}$) \\
		\hline
		Thin disc     & 0.23 & 0.036 \\
		Morgan disc   & 0.83 & 0.13 \\
	\end{tabular}
\end{table}

It is worth noting that the larger Morgan disc does not resolve the double-peak structure that is seen for the thin disc. The thin-disc model fit to the data is poor around 54000 MJD but the CRTS data (points prior to $\sim56000$ MJD) are less reliable at these fainter magnitudes. It is also possible that the single Sloan epoch of $g=23$ mag is not reliable and that the clear CRTS filter will introduce colour effects. These caveats aside, this lens model does provide a plausible explanation for the double-peaked structure seen in J142232 but requires refinement before any further conclusions can be drawn. As with J084305 and J094511, there is evidence to suggest that the BLR emission regions, particularly \ion{C}{iii]} but also \ion{Mg}{ii} in this case, are compact enough to show significant amplification changes on these time-scales.

\subsubsection{J103837}

In the case of J103837 there is even more complexity in the light curve. The CR lens example used for J142232 allows a comparison with double-peaked light curves, but to explain the three peaks seen for this object requires a more complex/realistic model. There is also spectroscopic evidence for an intervening system at $z=0.18$, which, if confirmed, would help reduce the model uncertainties. Assuming a lens mass of $0.1 {\rm M}_\odot$, this projects to an Einstein radius of $\sim8$ light-days. The thin-disc size estimate, based on the faintest of the three observed epochs is $\sim0.1$ light-days. The presence of an external shear with additional lens masses could provide an adequate explanation for the variability seen in this object. It will be the focus of future work.

\section{Discussion}
\label{sec:Discussion}

The AGN transients in this paper have been selected because they are candidates for microlensing events. The photometry shows a minimum of a factor 5 increase in luminosity and a smooth evolution on year-long time-scales. For three of the objects, the spectroscopy reveals differential evolution in the broad line fluxes with respect to the continuum. In particular, \ion{C}{iii]} is seen to track the changes in the continuum more closely than that of \ion{Mg}{ii}. Two objects also show spectroscopic signatures indicating the possible presence of an intervening galaxy. A simple point-source/point-lens model produces a reasonable fit to the bulk continuum changes seen for the single-peaked events. The testing of extended accretion-disc models with these lens configurations, given the parameter uncertainties, suggests the point-source approximation is reasonable but does not allow us to rule out the possibility that the accretion discs are being resolved, if only marginally. When extended source models are included for the BLR, the differential changes seen in the broad lines can be interpreted as the BLR being partially resolved by the lens. This implies that the \ion{C}{iii]} emitting region is smaller than that for \ion{Mg}{ii} and the data for J084305 allows size constraints for these regions of the order $\sim2\textup{--}11$ light-days. More complex lensing models are required to produce light curves displaying multiple peaks.

Having summarized our results so far, we now put them into context of what we know about AGN, discuss whether they make sense and what we might learn from them. The focus for this discussion will be on the microlensing hypothesis. Other plausible scenarios include the following: TDEs; accretion-disc instabilities; and extinction events. These are considered in more detail in L16. Target J142232 also has X-ray data available. \citet{Collinson2016} shows that, based on the broad-band SED, a lensing scenario is consistent with the data.

\subsection{Continuum variability}

Can the variability of the AGN in this paper be described as typical? In the context of the wider AGN population, long-term optical variability is commonly described in terms of a damped-random-walk (DRW) model \citep{Kelly2009,MacLeod2010,MacLeod2012}. It quantifies the variability using a structure function, which gives the rms magnitude difference as a function of time lag between different epochs. This model performs well when applied to large quasar samples. The two key parameters are the damping or characteristic time-scale, $\tau$, and the asymptotic rms magnitude difference at the longest time-scales, SF$_{\infty}$. We obtain the DRW parameters for J084305 and J094511 in two situations. First, for the original light curve and second, for the residual light curve after taking the microlensing model into account. The values obtained are displayed in Table \ref{tab:DRW}. 

\begin{table}
	\centering
	\caption{Damped-random-walk parameters obtained for J084305 and J094511. The values have been computed for the observed data and the residuals after subtraction of the microlensing model.}
	\label{tab:DRW}
	\begin{tabular}{lllll}
		& J084305 & & J094511 \\
		& Observed & Residual & Observed & Residual \\
		\hline
		SF$_{\infty}$     & 1.62 & 0.21 & 2.19 & 0.26 \\
		${\rm log}(\tau)$ & 3.72 & 1.96 & 3.94 & 2.28 \\
	\end{tabular}
\end{table}

With reference to \citet[Table 2]{MacLeod2010}, it is clear the DRW parameter values for the observed data are atypical for quasars and may even be biased due to a different underlying process describing the observed variability \citep{Kozlowski2016}. However, after taking the microlensing models into account, the parameters [${\rm log}(\tau)\sim2, {\rm SF}_\infty\sim 0.2$] are much more typical for quasars. A typical DRW origin for the bulk variability in these objects cannot be conclusively ruled out but this does lend further credence to the microlensing hypothesis. Regardless of the physical model it is notable that, after removing the trend, they look normal.

The underlying mechanism responsible for typical AGN variability is thought to be related to accretion-disc thermal instabilities and/or reprocessing of X-ray/UV emission \citep{Kelly2011,Shappee2014,Edelson2015}. There may also be reprocessing of FUV emission due to optically thick clouds on the inner edge of the BLR. This can reproduce some of the observed UV/optical lags \citep{Gardner2016} and can also address issues in relation to observations of the UV bump \citep{Lawrence2012}. In effect, these clouds would give rise to a `pseudo-continuum' component, which may prove important when testing various surface brightness profiles relating to the accretion disc and BLR. It has been suggested that the variability seen in luminous quasars on long time-scales may be primarily due to ongoing microlensing \citep{Hawkins2002}.

\subsection{Accretion disc: point-source or extended?}

Is the point-source/point-lens model an oversimplification? To first order, the model produces a reasonable fit to the bulk continuum changes in J084305/J094511 and also allows constraints on the lens parameters, which are reasonable given that there is currently no confirmed lens redshift. In addition, the testing of extended accretion discs with these lens parameters, using two different estimates for the true disc size, also produces reasonable fits to the data. The thin-disc prediction is broadly consistent but we can rule out the \citet{Morgan2010} for some lens configurations. Thus, we can be fairly confident in the validity of the point-source approximation but note that it is still possible that the disc is being resolved by the lens. Broadly speaking, for an accretion disc of approximately $\sim0.1\,\theta_{\rm E}$ or greater, it becomes possible to differentiate from the point-source solution using our data. Early further testing using MCMC models and extended sources shows tentative evidence that the disc is being resolved, but, given the large uncertainties on the Einstein radius of the lens, we are not yet able to place meaningful constraints on the absolute size of the accretion disc in these objects. In this work, we have also made the simplifying assumption that the accretion-disc surface brightness profile can be treated as a Gaussian, with a 2$\sigma$ radius equivalent to the radii calculated in Equations \ref{eq:thindisc} \& \ref{eq:Morgan}. This should prove reasonable as \citet{Mortonson2005} state that microlensing fluctuations are relatively insensitive to all circular disc model properties with the exception of the half-light radii.

The colour changes seen in the UV power-law slopes for these objects at later epochs (Table \ref{tab:data}) are an indication that the disc is at least partially resolved by the lens. These changes would be sensitive to the temperature profile of the disc and also the presence of any caustic structure in the lens configuration. For the reasons outlined above, we have not attempted to constrain the disc profile but hope to explore this in a follow-up paper. Alternatively, changes in amplification between the disc and a more extended `pseudo-continuum' cloud component could also produce the observed colour change though this possibility has not been fully explored \citep{Lawrence2012}. At later epochs, host/lens galaxy contamination may also need to be accounted for. For J084305/J094511, the estimate for the expected host/lens galaxy contribution is at least a factor of 2--3 below the AGN base level in the $g$ band. This goes some way in explaining why the detection of lens galaxy signatures in the spectra for these objects has proven difficult.

A further consideration is the effect of any broad line flux intruding into the filters used in the observations, particularly with respect to \ion{Mg}{ii} in the $g$ band. In the MCMC modelling for J084305/J094511, this extra flux component should be accounted for within the unlensed flux parameter, which is not currently well constrained, but should not invalidate the results. The spectra have also been scaled to the microlensing model, using an LT $g$-band transmission function to measure the flux, in order to keep the spectral measurements consistent with the photometry. A sign that the microlensing model for J094511 performs less well than expected is the resulting significant flux increase seen in the third epoch for \ion{Mg}{ii}/\ion{[O}{ii]}. The additional structure in the light curve at late epochs is the likely cause for this discrepancy. As the event continues to evolve/fade, it will be interesting to note whether or not this discrepancy is resolved with further modelling.

\subsection{BLR}

With the simplifying assumption that the variability in the broad lines is entirely due to lensing, the differential changes between the continuum, \ion{C}{iii]} and \ion{Mg}{ii} fluxes seen in three of the targets suggest that the BLR is at least partially resolved. Indeed, for J084305 the data is sufficient to allow constraints on the size of the \ion{C}{iii]} emitting region. These measurements are made with the assumption that our simplified lens model is appropriate but should necessarily be treated with a degree of caution. The size estimate from our model is $\sim5$ light-days with an expected range of $\sim1.7$--$11$ light-days when taking account the uncertainty in the Einstein radius of the lens. These sizes reflect the 2$\sigma$ radii used in our extended source modelling. Converting to half-light radii yields $\sim1.0$--$6.5$ light-days, which is surprisingly small when compared with BLR radius estimators. One example is from \citet{Bentz2009} for H$\,\beta$ and another from \citet{Kaspi2007} for \ion{C}{iv}. Using these, after obtaining a continuum luminosity for J084305 from the fourth epoch UV power-law and correcting for the lensing amplification, we obtain a radius of 34 and 31 light-days for H$\,\beta$ and \ion{C}{iv} respectively. Given these values, one would expect \ion{C}{iii]} to perhaps be at least comparable in size, if not a factor of up to 2 larger than that seen for \ion{C}{iv}. It is possible that the size estimate suffers due to the fact that our lens model is an oversimplification. This will be discussed in the next section.

For comparison with other observations, there are a handful of published lags for \ion{C}{iii]} that have been determined from reverberation mapping. These serve as a proxy for BLR radius and some notable results include \citet{Peterson1999,Onken2002,Metzroth2006} and \citet{Trevese2014}. Most are for low-redshift/luminosity targets and show \ion{C}{iii]} to be of the order of 3.5--30 light-days for these objects, with a considerable degree of uncertainty. The high-luminosity quasar in \citet{Trevese2014} has a much larger radius of the order of 270 light-days. For J084305, we estimate, after correcting for the lensing amplification, a bolometric luminosity for the AGN of ${\rm log}(L_{\rm bol})\sim44.7$. This is at the lower luminosity end of published lags but again suggests that our estimate is intriguingly small. Of particular interest is the result from \citet{Sluse2011}, where a radius for \ion{C}{iii]} has been estimated based on the microlensing of a multiply imaged quasar. Here, a multicomponent fit to the \ion{C}{iii]} profile provided evidence that the broadest components lie closer to the accretion disc and are therefore more affected by microlensing. In relative terms, their broadest component (for \ion{C}{iii]} and \ion{C}{iv}) is around four times larger than the continuum emitting region. The narrowest component is around a factor of 25 larger. Thus, our estimates for the \ion{C}{iii]} region in J084305, though small, may in fact be dominated by the changes in the inner component. Thus far, we have only used a single component for the \ion{C}{iii]} region, primarily due to SNR issues. An analysis of spectral ratios for J084305/J094511 shows tentative evidence for a changing very broad component redwards of the broad \ion{C}{iii]} line, though one must be careful not to over interpret the data. A further consideration is the true morphology of the \ion{C}{iii]} region. Reverberation mapping is sampling in the radial direction whereas microlensing provides a transverse sampling. Perhaps the \ion{C}{iii]} region is more compact along certain lines of sight than reverberation mapping implies.

For J084305, in contrast to \ion{C}{iii]}, the \ion{Mg}{ii} flux is consistent with no change. This implies that this region is more extended and therefore will not exhibit significant changes over the time-scale of our observations. The lower size limit obtained is based on a single component to the line. Limited testing with the use of a double Gaussian component to the line does not significantly affect the single component measurements or the Fe template fit. \citet{Sluse2012} show that both \ion{C}{iii]} and \ion{Mg}{ii} exhibit changes due to microlensing in a sample of multiply imaged quasars. They see evidence that the BLR is in general not spherically symmetric and that a bi-conical outflow model may not be appropriate in these cases. These systems benefit from the additional information supplied by the resolvable quasar images.

Thus far it has been assumed that the variability seen in the broad lines is entirely due to lensing but a more realistic scenario must include some intrinsic variability over the period of our observations. It may also be possible that, rather than being resolved by the lens, the observed broad line changes are entirely intrinsic to the AGN. Taking the \ion{Mg}{ii} line as an example, \citet{Sun2015} see significant variability in this line ($\sim10\%$) on $\sim100$ d time-scales. The amplitude of light-curve variability seen in \ion{Mg}{ii} also correlates with that seen in the continuum. Some reliable reverberation lags for \ion{Mg}{ii} are reported in \citet{Shen2016}, but a clear correlation with luminosity is not always apparent. In other cases, \ion{Mg}{ii} is unresponsive to continuum changes \citep{Cackett2015} but on longer time-scales \ion{Mg}{ii} does appear to respond to large changes in continuum flux \citep{MacLeod2016}. In the case of J084305 and J094511, the \ion{Mg}{ii} line does not seem to respond to the continuum drop even on these long time-scales. Given that the \ion{Mg}{ii} line should be at least partially dependent on the ionizing continuum, this is perhaps surprising. Lensing as an extrinsic cause for the continuum variability provides a natural solution however.

\subsection{Lensing system}
\label{sec:LensDiscuss}

For the two objects where the simple lensing model performs well, it is unfortunate that, as yet, there are no confirmed signatures of the presence of a lensing galaxy. This confirmation would allow us to move beyond the simple lensing models detailed here. At present, we are assuming that the convergence/shear (and hence macromagnification) are small enough to be neglected. Without confirmation of the presence of any multiple images or a reliable lens galaxy detection/redshift, the true lens parameters are very difficult to pin down and thus the results reported here should necessarily be treated with caution. However, there are a two points worth considering, particularly in the context of J084305/J094511, which help to put the assumption of a minimal convergence/shear on surer footing. Each will be discussed in turn.

First, the light curves are evolving smoothly. This is straightforward to do in the point-lens case but, in the presence of extended regions of caustic structure, one might expect more variation/asymmetry in the light curves. One solution is that the accretion disc is extended enough such that any caustic structure is poorly resolved. This could easily give rise to a single-peaked event without revealing the extra structure present in the magnification map. Preliminary testing of a more advanced MCMC approach has allowed us to make use of extended sources using a lower-resolution version of the CR magnification map ($\gamma=0.05$) used in the qualitative analysis for J142232. We add an eighth parameter for the source radius and the impact parameter is treated in the same way as for the point-lens case, i.e. an offset in the $y$-axis. Early results show that, with a good fit to the J084305 light curve, the most probable $2\sigma$ radius for the accretion disc is $\sim3$ light-days or $\sim0.2R_{\rm E}$. While, as expected, the effect of the additional shear was to increase the size of the accretion disc, this value is now larger than both empirical estimates for the disc size listed in Table \ref{tab:J084305ADsizes}.

Secondly, there is a large change in amplification of $\sim2\,{\rm mag}$ and the light curves are approximately symmetric. It is relatively straightforward to achieve light curves of this nature in low-kappa/gamma regimes (below $\sim 0.1$), as there will likely be isolated regions of high amplification in the magnification map. However, if the magnification map is a complex caustic network with many overlapping regions, it becomes more difficult to reproduce such light curves. As discussed in L16, a random foreground galaxy need not be particularly massive and it may be that these targets are not multiply imaged at all. Singly imaged sources are more likely to have an optical depth to microlensing below 0.1 than their multiply imaged counterparts \citep{Wyithe2001}.

Short of the confirmation of a lens galaxy, one way of improving the constraints on the lens parameters would be to include a prior on the transverse velocity of the lens in the MCMC analysis. For now, this parameter has been left free as a useful check on the validity of the results. The predicted values for the velocity are reasonable though perhaps a little high, particularly for J084305 with $v_\perp=830\,^{+1080}_{-510}$ km\,s$^{-1}$. If the lens is at a low redshift, then the transverse velocity for this target can be dominated by the peculiar velocity component of the lens galaxy, and values this high are not completely unreasonable \citep{Mosquera2011}. It also possible, though perhaps less likely, that some of these lensing events are due to isolated extragalactic point-masses.

\subsection{Future work}

One major drawback of this sample is that it is currently very difficult to remove any intrinsic AGN variability, preventing tighter constraints on the lensing parameters. This is possible for a number of multiply imaged quasars, where the time delays between images are short enough that intrinsic variations can be corrected for. For these, there is also a known lens galaxy. In contrast, our objects are among the lowest luminosity and lowest redshift AGN lensing candidates known. The SDSS $i$-band magnitudes for J084305/J094511 are 20.98/21.11 mag (DR12 cModelMag), respectively; comparison with \citet{Mosquera2011} shows that there are only four targets at $z<1$ and 10/87 with $i>20$ mag. Whilst it may not be possible to disentangle the intrinsic AGN variability from the observations, these low-luminosity targets do allow for a greater chance of observing larger amplitude changes in the BLR. In addition, the lensing models used for these objects may prove to be comparatively simple when compared to those involving massive elliptical lensing galaxies.

A priority for the future is to obtain high-resolution imaging of these targets in order to ascertain if there is evidence of any strong lensing or indeed the lens galaxy itself. There is also a good prospect of detecting many more lensed AGN candidates from long-baseline photometry. In addition, the slow evolution and high amplitude maximize the chances for obtaining follow-up observations of each event. For J084305/J094511, the models give an Einstein time of 7.5/10.8 yr and peak magnification factor for the continuum of 10.3/13.5, respectively. Long-term surveys should provide the opportunity to select targets on the slow rise to a high-amplitude microlensing event and allow ample time to plan follow-up observations accordingly. It would prove invaluable to monitor one of these microlensing events with a cadence that allows reverberation mapping to be performed simultaneously.

It is likely that more of the objects in our larger, highly variable AGN sample are also undergoing microlensing events. For some, the simple model again provides a good fit to the data though their slow evolution meant that they were not included in this paper. Other targets are excellent candidates for microlensing events in which the accretion disc may be well resolved by the lens. These, as with J142232/J103837, require a more thorough analysis in order to fully address these questions. More complex magnification maps, such as those available through the {\tt GERLUMPH} data base \citep{Vernardos2015} are also being considered. In addition, incorporating all available photometry bands into the analysis is a high priority for future work.

If the microlensing interpretation is correct, this provides a valuable way of probing the inner regions of AGN. This technique is already being used for strong-lensed, multiply imaged quasars \citep{Sluse2012,Sluse2015}. If the lens-host galaxy is small and the event is due to a single stellar lens, these objects will be in a different regime to their multiply imaged cousins. This, coupled with the fact that they are high-amplification events evolving over year-long time-scales, also allows a coordinated observational strategy. As the microlensing event unfolds, it can provide a transverse sampling of the BLR, perhaps even the accretion disc. The shorter time-scale variability allows a probe of the radial structure via reverberation mapping. These high-amplitude microlensing events may be unique opportunities for mapping the inner structure of an AGN.

\section{Conclusions}

In summary, we have analysed four extreme AGN transients, selected as candidates for rare, high-amplitude microlensing events. The light-curve information, primarily from the LT and supplemented by data from Pan-STARRS and SDSS has allowed a detailed MCMC analysis of a simple point-lens point-source model. This model has proven to be an excellent fit to the data in the two single-peaked cases, J084305 and J094511. After removing the signal due to microlensing, the long-term variability in these objects appears much more typical of the wider AGN population.

The multi-epoch spectroscopy from the WHT has also provided valuable insight into these events. All four targets display differential variability of the broad line fluxes with respect to the continuum. In three objects, the \ion{C}{iii]} flux is seen to track the continuum closely whereas the \ion{Mg}{ii} flux is less responsive to these changes or shows signs of no change at all. Using the working lensing models and the simplifying assumption of Gaussian surface brightness profiles, the changes can be interpreted as being due to regions of differing size being affected by the lensing event to a different extent. The continuum emission in the $g$ band can be reasonably treated as a point-source and the \ion{C}{iii]} region is more compact than for \ion{Mg}{ii}. In the case of J084305, the data also allows size constraints to be placed on these regions.

The two objects not well described by the simple lensing models are J142232 and J103837. In both cases, there is evidence for an intervening absorber, suggestive of the presence of a lensing galaxy. It is clear that the extra structure in the light curve for these double- or multiply peaked events, requires a more complex lensing scenario to describe the event. This will likely involve the presence of additional lens masses and/or external shear with the resulting caustic network giving rise to the features seen in the light curves. Any thorough analysis of these more complex models will also require consideration of extended sources, both for the accretion disc and BLR. Current limitations have meant that this will be left for future work, but microlensing is still the favoured scenario for the long-term variability seen in these objects.

Future time-domain surveys will be an invaluable source of microlensing events of this type. A challenge to overcome will be in identifying the gradual trends over a year or more that might indicate a high-amplitude event is imminent. If possible, this may allow sufficient time to coordinate an observing strategy, particularly with regard to reverberation mapping. The amplification, by an order of magnitude or more, of a higher redshift, lower luminosity AGN would be an ideal candidate for a Reverberation Mapping (RM) study. This, combined with a detailed lensing analysis, allows an unprecedented opportunity for probing the inner regions of these AGN. They are in a different regime, but wholly complimentary, to the microlensing studies in more luminous, multiply imaged AGN. We must make the most of these opportunities whenever they arise.

\section*{Acknowledgements}

\defcitealias{Astropy2013}{Astropy Collaboration, 2013}

AB acknowledges the support of the University of Edinburgh via the Principal's Career Development Scholarship. Thanks also to Marianne Vestergaard for providing the Fe(UV) template and to Jorge Jim\'enez-Vicente for supplying the base microlensing IRS code, which we adapted for use in the extended model testing. This research made use of Astropy, a community-developed core Python package for Astronomy \citepalias{Astropy2013}.

The Pan-STARRS1 Surveys (PS1) have been made possible through contributions of the Institute for Astronomy, the University of Hawaii, the Pan-STARRS Project Office, the Max-Planck Society and its participating institutes, the Max Planck Institute for Astronomy, Heidelberg and the Max Planck Institute for Extraterrestrial Physics, Garching, The Johns Hopkins University, Durham University, the University of Edinburgh, Queen's University Belfast, the Harvard-Smithsonian Center for Astrophysics, the Las Cumbres Observatory Global Telescope Network Incorporated, the National Central University of Taiwan, the Space Telescope Science Institute, the National Aeronautics and Space Administration under Grant no. NNX08AR22G issued through the Planetary Science Division of the NASA Science Mission Directorate, the National Science Foundation under Grant no. AST-1238877, the University of Maryland, and Eotvos Lorand University (ELTE) and the Los Alamos National Laboratory.

The Liverpool Telescope is operated on the island of La Palma by Liverpool John Moores University in the Spanish Observatorio del Roque de los Muchachos of the Instituto de Astrof\'isica de Canarias with financial support from the UK Science and Technology Facilities Council.

The William Herschel Telescope is operated on the island of La Palma by the Isaac Newton Group in the Spanish Observatorio del Roque de los Muchachos of the Instituto de Astrof\'isica de Canarias.

%%%%%%%%%%%%%%%%%%%%%%%%%%%%%%%%%%%%%%%%%%%%%%%%%%

%%%%%%%%%%%%%%%%%%%% REFERENCES %%%%%%%%%%%%%%%%%%

% The best way to enter references is to use BibTeX:

\bibliographystyle{mnras}
\bibliography{Bruce2016_ADS} % if your bibtex file is called example.bib

\begin{thebibliography}{}
\makeatletter
\relax
\def\mn@urlcharsother{\let\do\@makeother \do\$\do\&\do\#\do\^\do\_\do\%\do\~}
\def\mn@doi{\begingroup\mn@urlcharsother \@ifnextchar [ {\mn@doi@}
  {\mn@doi@[]}}
\def\mn@doi@[#1]#2{\def\@tempa{#1}\ifx\@tempa\@empty \href
  {http://dx.doi.org/#2} {doi:#2}\else \href {http://dx.doi.org/#2} {#1}\fi
  \endgroup}
\def\mn@eprint#1#2{\mn@eprint@#1:#2::\@nil}
\def\mn@eprint@arXiv#1{\href {http://arxiv.org/abs/#1} {{\tt arXiv:#1}}}
\def\mn@eprint@dblp#1{\href {http://dblp.uni-trier.de/rec/bibtex/#1.xml}
  {dblp:#1}}
\def\mn@eprint@#1:#2:#3:#4\@nil{\def\@tempa {#1}\def\@tempb {#2}\def\@tempc
  {#3}\ifx \@tempc \@empty \let \@tempc \@tempb \let \@tempb \@tempa \fi \ifx
  \@tempb \@empty \def\@tempb {arXiv}\fi \@ifundefined
  {mn@eprint@\@tempb}{\@tempb:\@tempc}{\expandafter \expandafter \csname
  mn@eprint@\@tempb\endcsname \expandafter{\@tempc}}}

\bibitem[\protect\citeauthoryear{{Ahn} et~al.,}{{Ahn} et~al.}{2012}]{Ahn2012}
{Ahn} C.~P.,  et~al., 2012, \mn@doi [\apjs] {10.1088/0067-0049/203/2/21}, \href
  {http://adsabs.harvard.edu/abs/2012ApJS..203...21A} {203, 21}

\bibitem[\protect\citeauthoryear{{Astropy Collaboration} et~al.,}{{Astropy
  Collaboration} et~al.}{2013}]{Astropy2013}
{Astropy Collaboration} et~al., 2013, \mn@doi [\aap]
  {10.1051/0004-6361/201322068}, \href
  {http://adsabs.harvard.edu/abs/2013A%26A...558A..33A} {558, A33}

\bibitem[\protect\citeauthoryear{{Bentz} et~al.,}{{Bentz}
  et~al.}{2009}]{Bentz2009}
{Bentz} M.~C.,  et~al., 2009, \mn@doi [\apj] {10.1088/0004-637X/705/1/199},
  \href {http://adsabs.harvard.edu/abs/2009ApJ...705..199B} {705, 199}

\bibitem[\protect\citeauthoryear{{Blackburne}, {Pooley}, {Rappaport}  \&
  {Schechter}}{{Blackburne} et~al.}{2011}]{Blackburne2011}
{Blackburne} J.~A.,  {Pooley} D.,  {Rappaport} S.,   {Schechter} P.~L.,  2011,
  \mn@doi [\apj] {10.1088/0004-637X/729/1/34}, \href
  {http://adsabs.harvard.edu/abs/2011ApJ...729...34B} {729, 34}

\bibitem[\protect\citeauthoryear{{Cackett}, {G{\"u}ltekin}, {Bentz},
  {Fausnaugh}, {Peterson}, {Troyer}  \& {Vestergaard}}{{Cackett}
  et~al.}{2015}]{Cackett2015}
{Cackett} E.~M.,  {G{\"u}ltekin} K.,  {Bentz} M.~C.,  {Fausnaugh} M.~M.,
  {Peterson} B.~M.,  {Troyer} J.,   {Vestergaard} M.,  2015, \mn@doi [\apj]
  {10.1088/0004-637X/810/2/86}, \href
  {http://adsabs.harvard.edu/abs/2015ApJ...810...86C} {810, 86}

\bibitem[\protect\citeauthoryear{{Cardelli}, {Clayton}  \& {Mathis}}{{Cardelli}
  et~al.}{1989}]{Cardelli1989}
{Cardelli} J.~A.,  {Clayton} G.~C.,   {Mathis} J.~S.,  1989, \mn@doi [\apj]
  {10.1086/167900}, \href {http://adsabs.harvard.edu/abs/1989ApJ...345..245C}
  {345, 245}

\bibitem[\protect\citeauthoryear{{Chabrier}}{{Chabrier}}{2003}]{Chabrier2003}
{Chabrier} G.,  2003, \mn@doi [\pasp] {10.1086/376392}, \href
  {http://adsabs.harvard.edu/abs/2003PASP..115..763C} {115, 763}

\bibitem[\protect\citeauthoryear{{Chang} \& {Refsdal}}{{Chang} \&
  {Refsdal}}{1984}]{Chang1984}
{Chang} K.,  {Refsdal} S.,  1984, \aap, \href
  {http://adsabs.harvard.edu/abs/1984A%26A...132..168C} {132, 168}

\bibitem[\protect\citeauthoryear{{Collinson}}{{Collinson}}{2016}]{Collinson2016}
{Collinson} J.,  2016, \mnras, submitted

\bibitem[\protect\citeauthoryear{{Drake} et~al.,}{{Drake}
  et~al.}{2009}]{Drake2009}
{Drake} A.~J.,  et~al., 2009, \mn@doi [\apj] {10.1088/0004-637X/696/1/870},
  \href {http://adsabs.harvard.edu/abs/2009ApJ...696..870D} {696, 870}

\bibitem[\protect\citeauthoryear{{Edelson} et~al.,}{{Edelson}
  et~al.}{2015}]{Edelson2015}
{Edelson} R.,  et~al., 2015, \mn@doi [\apj] {10.1088/0004-637X/806/1/129},
  \href {http://adsabs.harvard.edu/abs/2015ApJ...806..129E} {806, 129}

\bibitem[\protect\citeauthoryear{{Eigenbrod}, {Courbin}, {Sluse}, {Meylan}  \&
  {Agol}}{{Eigenbrod} et~al.}{2008}]{Eigenbrod2008}
{Eigenbrod} A.,  {Courbin} F.,  {Sluse} D.,  {Meylan} G.,   {Agol} E.,  2008,
  \mn@doi [\aap] {10.1051/0004-6361:20078703}, \href
  {http://adsabs.harvard.edu/abs/2008A%26A...480..647E} {480, 647}

\bibitem[\protect\citeauthoryear{{Elvis} et~al.,}{{Elvis}
  et~al.}{1994}]{Elvis1994}
{Elvis} M.,  et~al., 1994, \mn@doi [\apjs] {10.1086/192093}, \href
  {http://adsabs.harvard.edu/abs/1994ApJS...95....1E} {95, 1}

\bibitem[\protect\citeauthoryear{{Foreman-Mackey}, {Hogg}, {Lang}  \&
  {Goodman}}{{Foreman-Mackey} et~al.}{2013}]{Foreman-Mackey2013}
{Foreman-Mackey} D.,  {Hogg} D.~W.,  {Lang} D.,   {Goodman} J.,  2013, \mn@doi
  [\pasp] {10.1086/670067}, \href
  {http://adsabs.harvard.edu/abs/2013PASP..125..306F} {125, 306}

\bibitem[\protect\citeauthoryear{{Gardner} \& {Done}}{{Gardner} \&
  {Done}}{2016}]{Gardner2016}
{Gardner} E.,  {Done} C.,  2016, preprint, \href
  {http://adsabs.harvard.edu/abs/2016arXiv160309564G} {} (\mn@eprint {arXiv}
  {1603.09564})

\bibitem[\protect\citeauthoryear{{Hawkins}}{{Hawkins}}{2002}]{Hawkins2002}
{Hawkins} M.~R.~S.,  2002, \mn@doi [\mnras] {10.1046/j.1365-8711.2002.04939.x},
  \href {http://adsabs.harvard.edu/abs/2002MNRAS.329...76H} {329, 76}

\bibitem[\protect\citeauthoryear{{Irwin}, {Webster}, {Hewett}, {Corrigan}  \&
  {Jedrzejewski}}{{Irwin} et~al.}{1989}]{Irwin1989}
{Irwin} M.~J.,  {Webster} R.~L.,  {Hewett} P.~C.,  {Corrigan} R.~T.,
  {Jedrzejewski} R.~I.,  1989, \mn@doi [\aj] {10.1086/115272}, \href
  {http://adsabs.harvard.edu/abs/1989AJ.....98.1989I} {98, 1989}

\bibitem[\protect\citeauthoryear{{Jim{\'e}nez-Vicente}}{{Jim{\'e}nez-Vicente}}{2016}]{Jimenez-Vicente2016}
{Jim{\'e}nez-Vicente} J.,  2016, {Astrophysical Applications of Gravitational
  Lensing}.
Cambridge University Press, Cambridge, p. 251

\bibitem[\protect\citeauthoryear{{Jim{\'e}nez-Vicente}, {Mediavilla},
  {Mu{\~n}oz}  \& {Kochanek}}{{Jim{\'e}nez-Vicente}
  et~al.}{2012}]{Jimenez-Vicente2012}
{Jim{\'e}nez-Vicente} J.,  {Mediavilla} E.,  {Mu{\~n}oz} J.~A.,   {Kochanek}
  C.~S.,  2012, \mn@doi [\apj] {10.1088/0004-637X/751/2/106}, \href
  {http://adsabs.harvard.edu/abs/2012ApJ...751..106J} {751, 106}

\bibitem[\protect\citeauthoryear{{Jim{\'e}nez-Vicente}, {Mediavilla},
  {Kochanek}, {Mu{\~n}oz}, {Motta}, {Falco}  \&
  {Mosquera}}{{Jim{\'e}nez-Vicente} et~al.}{2014}]{Jimenez-Vicente2014}
{Jim{\'e}nez-Vicente} J.,  {Mediavilla} E.,  {Kochanek} C.~S.,  {Mu{\~n}oz}
  J.~A.,  {Motta} V.,  {Falco} E.,   {Mosquera} A.~M.,  2014, \mn@doi [\apj]
  {10.1088/0004-637X/783/1/47}, \href
  {http://adsabs.harvard.edu/abs/2014ApJ...783...47J} {783, 47}

\bibitem[\protect\citeauthoryear{{Kaspi}, {Brandt}, {Maoz}, {Netzer},
  {Schneider}  \& {Shemmer}}{{Kaspi} et~al.}{2007}]{Kaspi2007}
{Kaspi} S.,  {Brandt} W.~N.,  {Maoz} D.,  {Netzer} H.,  {Schneider} D.~P.,
  {Shemmer} O.,  2007, \mn@doi [\apj] {10.1086/512094}, \href
  {http://adsabs.harvard.edu/abs/2007ApJ...659..997K} {659, 997}

\bibitem[\protect\citeauthoryear{{Kelly}, {Bechtold}  \&
  {Siemiginowska}}{{Kelly} et~al.}{2009}]{Kelly2009}
{Kelly} B.~C.,  {Bechtold} J.,   {Siemiginowska} A.,  2009, \mn@doi [\apj]
  {10.1088/0004-637X/698/1/895}, \href
  {http://adsabs.harvard.edu/abs/2009ApJ...698..895K} {698, 895}

\bibitem[\protect\citeauthoryear{{Kelly}, {Sobolewska}  \&
  {Siemiginowska}}{{Kelly} et~al.}{2011}]{Kelly2011}
{Kelly} B.~C.,  {Sobolewska} M.,   {Siemiginowska} A.,  2011, \mn@doi [\apj]
  {10.1088/0004-637X/730/1/52}, \href
  {http://adsabs.harvard.edu/abs/2011ApJ...730...52K} {730, 52}

\bibitem[\protect\citeauthoryear{{Koz{\l}owski}}{{Koz{\l}owski}}{2016}]{Kozlowski2016}
{Koz{\l}owski} S.,  2016, \mn@doi [\mnras] {10.1093/mnras/stw819}, \href
  {http://adsabs.harvard.edu/abs/2016MNRAS.459.2787K} {459, 2787}

\bibitem[\protect\citeauthoryear{{Lawrence}}{{Lawrence}}{2012}]{Lawrence2012}
{Lawrence} A.,  2012, \mn@doi [\mnras] {10.1111/j.1365-2966.2012.20889.x},
  \href {http://adsabs.harvard.edu/abs/2012MNRAS.423..451L} {423, 451}

\bibitem[\protect\citeauthoryear{{Lawrence} et~al.,}{{Lawrence}
  et~al.}{2016}]{Lawrence2016}
{Lawrence} A.,  et~al., 2016, \mn@doi [\mnras] {10.1093/mnras/stw1963}, \href
  {http://adsabs.harvard.edu/abs/2016MNRAS.463..296L} {463, 296}

\bibitem[\protect\citeauthoryear{{Lewis} \& {Irwin}}{{Lewis} \&
  {Irwin}}{1995}]{Lewis1995}
{Lewis} G.~F.,  {Irwin} M.~J.,  1995, \mn@doi [\mnras]
  {10.1093/mnras/276.1.103}, \href
  {http://adsabs.harvard.edu/abs/1995MNRAS.276..103L} {276, 103}

\bibitem[\protect\citeauthoryear{{MacLeod} et~al.,}{{MacLeod}
  et~al.}{2010}]{MacLeod2010}
{MacLeod} C.~L.,  et~al., 2010, \mn@doi [\apj] {10.1088/0004-637X/721/2/1014},
  \href {http://adsabs.harvard.edu/abs/2010ApJ...721.1014M} {721, 1014}

\bibitem[\protect\citeauthoryear{{MacLeod} et~al.,}{{MacLeod}
  et~al.}{2012}]{MacLeod2012}
{MacLeod} C.~L.,  et~al., 2012, \mn@doi [\apj] {10.1088/0004-637X/753/2/106},
  \href {http://adsabs.harvard.edu/abs/2012ApJ...753..106M} {753, 106}

\bibitem[\protect\citeauthoryear{{MacLeod} et~al.,}{{MacLeod}
  et~al.}{2015}]{MacLeod2015}
{MacLeod} C.~L.,  et~al., 2015, \mn@doi [\apj] {10.1088/0004-637X/806/2/258},
  \href {http://adsabs.harvard.edu/abs/2015ApJ...806..258M} {806, 258}

\bibitem[\protect\citeauthoryear{{MacLeod} et~al.,}{{MacLeod}
  et~al.}{2016}]{MacLeod2016}
{MacLeod} C.~L.,  et~al., 2016, \mn@doi [\mnras] {10.1093/mnras/stv2997}, \href
  {http://adsabs.harvard.edu/abs/2016MNRAS.457..389M} {457, 389}

\bibitem[\protect\citeauthoryear{{Magnier} et~al.,}{{Magnier}
  et~al.}{2013}]{Magnier2013}
{Magnier} E.~A.,  et~al., 2013, \mn@doi [\apjs] {10.1088/0067-0049/205/2/20},
  \href {http://adsabs.harvard.edu/abs/2013ApJS..205...20M} {205, 20}

\bibitem[\protect\citeauthoryear{{McLure} \& {Dunlop}}{{McLure} \&
  {Dunlop}}{2004}]{McLure2004}
{McLure} R.~J.,  {Dunlop} J.~S.,  2004, \mn@doi [\mnras]
  {10.1111/j.1365-2966.2004.08034.x}, \href
  {http://adsabs.harvard.edu/abs/2004MNRAS.352.1390M} {352, 1390}

\bibitem[\protect\citeauthoryear{{Metzroth}, {Onken}  \& {Peterson}}{{Metzroth}
  et~al.}{2006}]{Metzroth2006}
{Metzroth} K.~G.,  {Onken} C.~A.,   {Peterson} B.~M.,  2006, \mn@doi [\apj]
  {10.1086/505525}, \href {http://adsabs.harvard.edu/abs/2006ApJ...647..901M}
  {647, 901}

\bibitem[\protect\citeauthoryear{{Morgan}, {Kochanek}, {Morgan}  \&
  {Falco}}{{Morgan} et~al.}{2010}]{Morgan2010}
{Morgan} C.~W.,  {Kochanek} C.~S.,  {Morgan} N.~D.,   {Falco} E.~E.,  2010,
  \mn@doi [\apj] {10.1088/0004-637X/712/2/1129}, \href
  {http://adsabs.harvard.edu/abs/2010ApJ...712.1129M} {712, 1129}

\bibitem[\protect\citeauthoryear{{Mortonson}, {Schechter}  \&
  {Wambsganss}}{{Mortonson} et~al.}{2005}]{Mortonson2005}
{Mortonson} M.~J.,  {Schechter} P.~L.,   {Wambsganss} J.,  2005, \mn@doi [\apj]
  {10.1086/431195}, \href {http://adsabs.harvard.edu/abs/2005ApJ...628..594M}
  {628, 594}

\bibitem[\protect\citeauthoryear{{Mosquera} \& {Kochanek}}{{Mosquera} \&
  {Kochanek}}{2011}]{Mosquera2011}
{Mosquera} A.~M.,  {Kochanek} C.~S.,  2011, \mn@doi [\apj]
  {10.1088/0004-637X/738/1/96}, \href
  {http://adsabs.harvard.edu/abs/2011ApJ...738...96M} {738, 96}

\bibitem[\protect\citeauthoryear{{Onken} \& {Peterson}}{{Onken} \&
  {Peterson}}{2002}]{Onken2002}
{Onken} C.~A.,  {Peterson} B.~M.,  2002, \mn@doi [\apj] {10.1086/340351}, \href
  {http://adsabs.harvard.edu/abs/2002ApJ...572..746O} {572, 746}

\bibitem[\protect\citeauthoryear{{Peterson} \& {Wandel}}{{Peterson} \&
  {Wandel}}{1999}]{Peterson1999}
{Peterson} B.~M.,  {Wandel} A.,  1999, \mn@doi [\apjl] {10.1086/312190}, \href
  {http://adsabs.harvard.edu/abs/1999ApJ...521L..95P} {521, L95}

\bibitem[\protect\citeauthoryear{{Peterson} et~al.,}{{Peterson}
  et~al.}{2013}]{Peterson2013}
{Peterson} B.~M.,  et~al., 2013, \mn@doi [\apj] {10.1088/0004-637X/779/2/109},
  \href {http://adsabs.harvard.edu/abs/2013ApJ...779..109P} {779, 109}

\bibitem[\protect\citeauthoryear{{Planck Collaboration XVI,}}{{Planck
  Collaboration XVI,}}{2014}]{Planck2014}
{Planck Collaboration XVI,} 2014, \mn@doi [\aap] {10.1051/0004-6361/201321591},
  \href {http://adsabs.harvard.edu/abs/2014A%26A...571A..16P} {571, A16}

\bibitem[\protect\citeauthoryear{{Schlafly} \& {Finkbeiner}}{{Schlafly} \&
  {Finkbeiner}}{2011}]{Schlafly2011}
{Schlafly} E.~F.,  {Finkbeiner} D.~P.,  2011, \mn@doi [\apj]
  {10.1088/0004-637X/737/2/103}, \href
  {http://adsabs.harvard.edu/abs/2011ApJ...737..103S} {737, 103}

\bibitem[\protect\citeauthoryear{{Schlafly} et~al.,}{{Schlafly}
  et~al.}{2012}]{Schlafly2012}
{Schlafly} E.~F.,  et~al., 2012, \mn@doi [\apj] {10.1088/0004-637X/756/2/158},
  \href {http://adsabs.harvard.edu/abs/2012ApJ...756..158S} {756, 158}

\bibitem[\protect\citeauthoryear{{Shappee} et~al.,}{{Shappee}
  et~al.}{2014}]{Shappee2014}
{Shappee} B.~J.,  et~al., 2014, \mn@doi [\apj] {10.1088/0004-637X/788/1/48},
  \href {http://adsabs.harvard.edu/abs/2014ApJ...788...48S} {788, 48}

\bibitem[\protect\citeauthoryear{{Shen} et~al.,}{{Shen}
  et~al.}{2016}]{Shen2016}
{Shen} Y.,  et~al., 2016, \mn@doi [\apj] {10.3847/0004-637X/818/1/30}, \href
  {http://adsabs.harvard.edu/abs/2016ApJ...818...30S} {818, 30}

\bibitem[\protect\citeauthoryear{{Sluse} et~al.,}{{Sluse}
  et~al.}{2011}]{Sluse2011}
{Sluse} D.,  et~al., 2011, \mn@doi [\aap] {10.1051/0004-6361/201016110}, \href
  {http://adsabs.harvard.edu/abs/2011A%26A...528A.100S} {528, A100}

\bibitem[\protect\citeauthoryear{{Sluse}, {Hutsem{\'e}kers}, {Courbin},
  {Meylan}  \& {Wambsganss}}{{Sluse} et~al.}{2012}]{Sluse2012}
{Sluse} D.,  {Hutsem{\'e}kers} D.,  {Courbin} F.,  {Meylan} G.,   {Wambsganss}
  J.,  2012, \mn@doi [\aap] {10.1051/0004-6361/201219125}, \href
  {http://adsabs.harvard.edu/abs/2012A%26A...544A..62S} {544, A62}

\bibitem[\protect\citeauthoryear{{Sluse}, {Hutsem{\'e}kers}, {Anguita},
  {Braibant}  \& {Riaud}}{{Sluse} et~al.}{2015}]{Sluse2015}
{Sluse} D.,  {Hutsem{\'e}kers} D.,  {Anguita} T.,  {Braibant} L.,   {Riaud} P.,
   2015, \mn@doi [\aap] {10.1051/0004-6361/201526832}, \href
  {http://adsabs.harvard.edu/abs/2015A%26A...582A.109S} {582, A109}

\bibitem[\protect\citeauthoryear{{Steele} et~al.,}{{Steele}
  et~al.}{2004}]{Steele2004}
{Steele} I.~A.,  et~al., 2004, in {Oschmann} Jr. J.~M.,  ed.,  \procspie Conf.
  Ser. Vol. 5489, Ground-based Telescopes. SPIE, Bellingham. p.~679

\bibitem[\protect\citeauthoryear{{Sun} et~al.,}{{Sun} et~al.}{2015}]{Sun2015}
{Sun} M.,  et~al., 2015, \mn@doi [\apj] {10.1088/0004-637X/811/1/42}, \href
  {http://adsabs.harvard.edu/abs/2015ApJ...811...42S} {811, 42}

\bibitem[\protect\citeauthoryear{{Tonry} et~al.,}{{Tonry}
  et~al.}{2012}]{Tonry2012}
{Tonry} J.~L.,  et~al., 2012, \mn@doi [\apj] {10.1088/0004-637X/750/2/99},
  \href {http://adsabs.harvard.edu/abs/2012ApJ...750...99T} {750, 99}

\bibitem[\protect\citeauthoryear{{Trevese}, {Perna}, {Vagnetti}, {Saturni}  \&
  {Dadina}}{{Trevese} et~al.}{2014}]{Trevese2014}
{Trevese} D.,  {Perna} M.,  {Vagnetti} F.,  {Saturni} F.~G.,   {Dadina} M.,
  2014, \mn@doi [\apj] {10.1088/0004-637X/795/2/164}, \href
  {http://adsabs.harvard.edu/abs/2014ApJ...795..164T} {795, 164}

\bibitem[\protect\citeauthoryear{{Vernardos} \& {Fluke}}{{Vernardos} \&
  {Fluke}}{2013}]{Vernardos2013}
{Vernardos} G.,  {Fluke} C.~J.,  2013, \mn@doi [\mnras]
  {10.1093/mnras/stt1076}, \href
  {http://adsabs.harvard.edu/abs/2013MNRAS.434..832V} {434, 832}

\bibitem[\protect\citeauthoryear{{Vernardos}, {Fluke}, {Bate}, {Croton}  \&
  {Vohl}}{{Vernardos} et~al.}{2015}]{Vernardos2015}
{Vernardos} G.,  {Fluke} C.~J.,  {Bate} N.~F.,  {Croton} D.,   {Vohl} D.,
  2015, \mn@doi [\apjs] {10.1088/0067-0049/217/2/23}, \href
  {http://adsabs.harvard.edu/abs/2015ApJS..217...23V} {217, 23}

\bibitem[\protect\citeauthoryear{{V{\'e}ron-Cetty}, {Joly}  \&
  {V{\'e}ron}}{{V{\'e}ron-Cetty} et~al.}{2004}]{Veron-Cetty2004}
{V{\'e}ron-Cetty} M.-P.,  {Joly} M.,   {V{\'e}ron} P.,  2004, \mn@doi [\aap]
  {10.1051/0004-6361:20035714}, \href
  {http://adsabs.harvard.edu/abs/2004A%26A...417..515V} {417, 515}

\bibitem[\protect\citeauthoryear{{Vestergaard} \& {Wilkes}}{{Vestergaard} \&
  {Wilkes}}{2001}]{Vestergaard2001}
{Vestergaard} M.,  {Wilkes} B.~J.,  2001, \mn@doi [\apjs] {10.1086/320357},
  \href {http://adsabs.harvard.edu/abs/2001ApJS..134....1V} {134, 1}

\bibitem[\protect\citeauthoryear{{Wambsganss}}{{Wambsganss}}{1992}]{Wambsganss1992}
{Wambsganss} J.,  1992, \mn@doi [\apj] {10.1086/170987}, \href
  {http://adsabs.harvard.edu/abs/1992ApJ...386...19W} {386, 19}

\bibitem[\protect\citeauthoryear{{Wambsganss}}{{Wambsganss}}{2006}]{Wambsganss2006}
{Wambsganss} J.,  2006, in {Meylan} G.,  {Jetzer} P.,  {North} P.,  {Schneider}
  P.,  {Kochanek} C.~S.,   {Wambsganss} J.,  eds, Saas-Fee Advanced Course 33:
  Gravitational Lensing: Strong, Weak and Micro. Springer, Berlin. p.~453

\bibitem[\protect\citeauthoryear{{Wyithe} \& {Turner}}{{Wyithe} \&
  {Turner}}{2002}]{Wyithe2001}
{Wyithe} J.~S.~B.,  {Turner} E.~L.,  2002, \mn@doi [\apj] {10.1086/338426},
  \href {http://adsabs.harvard.edu/abs/2002ApJ...567...18W} {567, 18}

\bibitem[\protect\citeauthoryear{{York} et~al.,}{{York}
  et~al.}{2006}]{York2006}
{York} D.~G.,  et~al., 2006, \mn@doi [\mnras]
  {10.1111/j.1365-2966.2005.10018.x}, \href
  {http://adsabs.harvard.edu/abs/2006MNRAS.367..945Y} {367, 945}

\bibitem[\protect\citeauthoryear{{van Dokkum}}{{van
  Dokkum}}{2001}]{VanDokkum2001}
{van Dokkum} P.~G.,  2001, \mn@doi [\pasp] {10.1086/323894}, \href
  {http://adsabs.harvard.edu/abs/2001PASP..113.1420V} {113, 1420}

\makeatother
\end{thebibliography}

% Alternatively you could enter them by hand, like this:
% This method is tedious and prone to error if you have lots of references
%\begin{thebibliography}{99}
%\bibitem[\protect\citeauthoryear{Author}{2012}]{Author2012}
%Author A.~N., 2013, Journal of Improbable Astronomy, 1, 1
%\bibitem[\protect\citeauthoryear{Others}{2013}]{Others2013}
%Others S., 2012, Journal of Interesting Stuff, 17, 198
%\end{thebibliography}

%%%%%%%%%%%%%%%%%%%%%%%%%%%%%%%%%%%%%%%%%%%%%%%%%%

%%%%%%%%%%%%%%%%% APPENDICES %%%%%%%%%%%%%%%%%%%%%

\appendix

\section{Appendix}

\begin{figure*}
	\includegraphics[width=2\columnwidth]{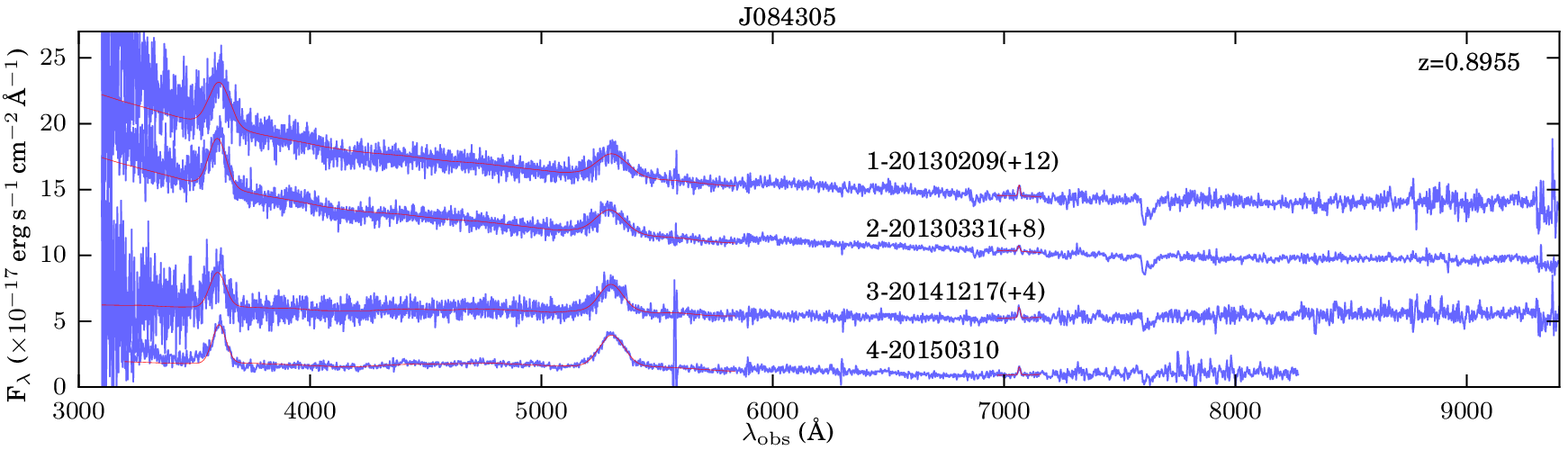}
    \caption{Spectra for J084305 as produced by our reduction pipeline. They have been scaled to our microlensing model (\ref{sec:MCMCresults}) and are corrected for Milky Way reddening. The red lines show the best-fitting models from the fitting process.}
    \label{fig:J084305_fit}
\end{figure*}

\begin{figure*}
	\includegraphics[width=2\columnwidth]{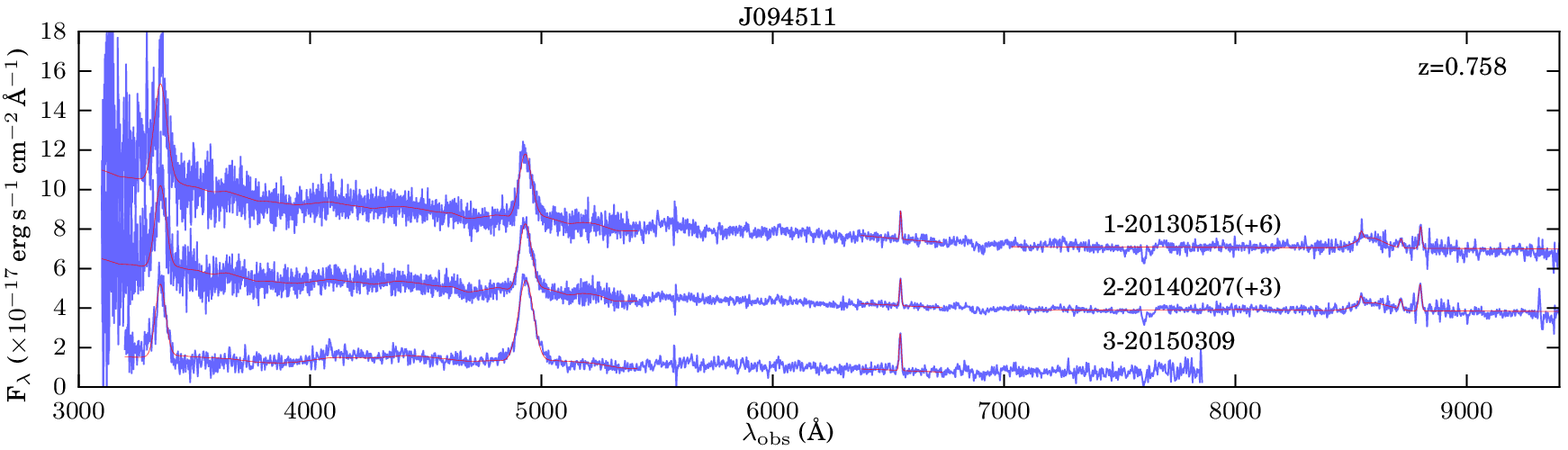}
    \caption{Spectra for J094511 as produced by our reduction pipeline. They have been scaled to our microlensing model (\ref{sec:MCMCresults}) and are corrected for Milky Way reddening. The red lines show the best-fitting models from the fitting process.}
    \label{fig:J094511_fit}
\end{figure*}

\begin{figure*}
	\includegraphics[width=2\columnwidth]{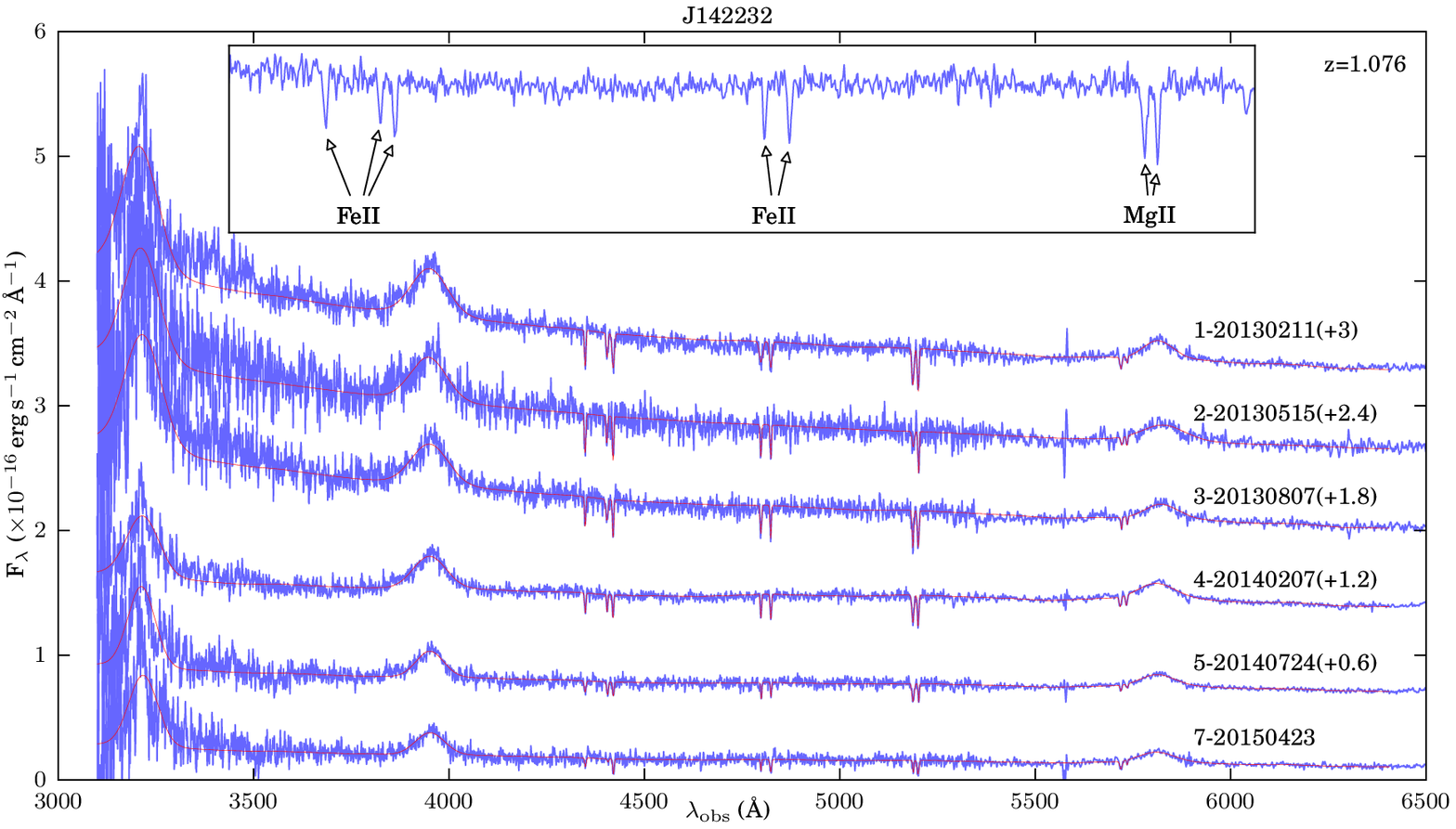}
    \caption{Spectra for J142232 as produced by our reduction pipeline. They have been scaled to our LT data (\ref{sec:Pipeline}) and are corrected for Milky Way reddening. The red lines show the best-fitting models from the fitting process. Inset is a zoomed region of the central portion of the fourth epoch to highlight the absorption complex.}
    \label{fig:J142232_fit}
\end{figure*}

\begin{figure*}
	\includegraphics[width=2\columnwidth]{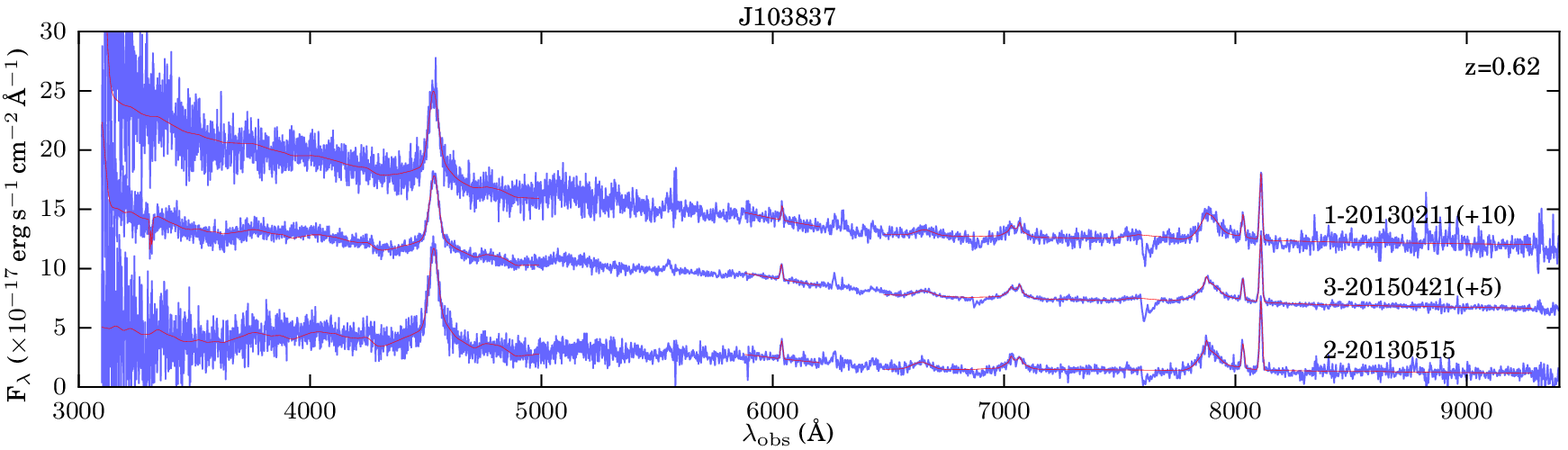}
    \caption{Spectra for J103837 as produced by our reduction pipeline. They have been scaled to our LT data (\ref{sec:Pipeline}) and are corrected for Milky Way reddening. The red lines show the best-fitting models from the fitting process.}
    \label{fig:J103837_fit}
\end{figure*}

%table updated 26/01/2016
\begin{table*}
	\centering
	\caption{Data table. (1) Scaling correction factor applied to the spetrum as described in \ref{sec:Pipeline}. (2) Power law slope from the fit given by Eq. \ref{eq:powerlaw}. (3) Observed AGN monochromatic continuum flux at rest-frame 3000\AA\, calculated from the power-law fit to the UV region. (4) AGN monochromatic continuum luminosity derived from (2). (5)/(6)/(7) Line centre, sigma and total line flux from the Gaussian fit to \ion{Mg}{ii} in the observed frame. For J103837 the values are from the narrower of the two broad \ion{Mg}{ii} components. (8) Black hole mass calculated using the \citet{McLure2004} relation. Errors are from the fitting process only.}
	\label{tab:data}
	\begin{tabular}{rrrrrrrrrr}
		\hline
		Target   & MJD & Scale & $\beta$ & $f_{3000}$ & $L_{3000}$ & $\lambdaup_\textrm{\ion{Mg}{ii}}$ & $\sigma_\textrm{\ion{Mg}{ii}}$ & $A_\textrm{\ion{Mg}{ii}}$ & $M_\textrm{BH}$ \\
		       & & & & $\scriptstyle\textrm{erg\,s}^{-1}\textrm{cm}^{-2}\textrm{\AA}^{-1}$ & $\scriptstyle\textrm{log$_{10}$(erg}\,\textrm{s}^{-1})$ & $\scriptstyle\textrm{\AA}$ & $\scriptstyle\textrm{\AA}$ & $\scriptstyle\textrm{erg}\,\textrm{s}^{-1}\,\textrm{cm}^{-2}$ & $\scriptstyle\textrm{log}_{10}(\textrm{M}_\odot)$ \\
		       &     &   (1) &  (2)       &    $\scriptstyle\times10^{-17}$ (3)    &  (4) &   (5)                   &       (6)              &    $\scriptstyle\times10^{-15}$ (7)     & (8)      \\
		\hline
J084305-1 & 56333.2 & 1.04 & -1.82$\pm$0.02 & 3.378$\pm$0.026 & 44.907$\pm$0.002 & 5304.9$\pm$2.2 & 63.0$\pm$2.3 & 2.84$\pm$0.09 & 8.91$\pm$0.03 \\
J084305-2 & 56382.9 & 0.83 & -1.86$\pm$0.01 & 3.035$\pm$0.018 & 44.861$\pm$0.002 & 5294.4$\pm$1.7 & 59.2$\pm$1.7 & 2.83$\pm$0.07 & 8.83$\pm$0.02 \\
J084305-3 & 57009.0 & 1.25 & -0.78$\pm$0.04 & 1.368$\pm$0.024 & 44.515$\pm$0.005 & 5299.5$\pm$1.5 & 51.0$\pm$1.5 & 2.91$\pm$0.07 & 8.49$\pm$0.02 \\
J084305-4 & 57092.2 & 1.42 & -0.82$\pm$0.05 & 1.099$\pm$0.029 & 44.420$\pm$0.008 & 5303.5$\pm$0.9 & 54.8$\pm$0.9 & 3.36$\pm$0.05 & 8.49$\pm$0.01 \\
J094511-1 & 56427.9 & 0.91 & -1.75$\pm$0.04 & 1.899$\pm$0.036 & 44.446$\pm$0.006 & 4930.3$\pm$0.6 & 28.5$\pm$0.6 & 2.60$\pm$0.05 & 8.00$\pm$0.02 \\
J094511-2 & 56695.9 & 0.80 & -1.75$\pm$0.03 & 1.308$\pm$0.023 & 44.284$\pm$0.005 & 4930.3$\pm$0.4 & 27.9$\pm$0.4 & 2.64$\pm$0.03 & 7.89$\pm$0.01 \\
J094511-3 & 57091.2 & 1.28 & -0.99$\pm$0.07 & 0.848$\pm$0.034 & 44.096$\pm$0.012 & 4932.9$\pm$0.4 & 32.9$\pm$0.4 & 3.50$\pm$0.04 & 7.91$\pm$0.01 \\
J142232-1 & 56335.2 & 1.04 & -1.90$\pm$0.01 & 3.008$\pm$0.017 & 45.093$\pm$0.002 & 5813.2$\pm$0.9 & 40.3$\pm$0.9 & 1.78$\pm$0.03 & 8.56$\pm$0.02 \\
J142232-2 & 56428.1 & 1.17 & -1.91$\pm$0.02 & 2.664$\pm$0.028 & 45.040$\pm$0.003 & 5825.9$\pm$1.9 & 55.1$\pm$2.2 & 1.93$\pm$0.06 & 8.80$\pm$0.03 \\
J142232-3 & 56511.9 & 1.02 & -1.99$\pm$0.02 & 2.238$\pm$0.018 & 44.964$\pm$0.002 & 5820.3$\pm$1.3 & 44.8$\pm$1.3 & 1.69$\pm$0.04 & 8.57$\pm$0.02 \\
J142232-4 & 56696.2 & 0.80 & -1.13$\pm$0.01 & 1.892$\pm$0.011 & 44.891$\pm$0.002 & 5807.5$\pm$0.7 & 49.3$\pm$0.8 & 2.02$\pm$0.02 & 8.61$\pm$0.01 \\
J142232-5 & 56862.9 & 0.89 & -1.43$\pm$0.02 & 1.116$\pm$0.010 & 44.662$\pm$0.003 & 5810.9$\pm$0.7 & 46.3$\pm$0.8 & 1.37$\pm$0.02 & 8.42$\pm$0.01 \\
J142232-7 & 57136.1 & 0.77 & -1.32$\pm$0.02 & 1.061$\pm$0.012 & 44.640$\pm$0.003 & 5808.5$\pm$1.1 & 48.5$\pm$1.2 & 1.25$\pm$0.02 & 8.44$\pm$0.02 \\
J103837-1 & 56335.1 & 1.42 & -1.92$\pm$0.03 & 6.016$\pm$0.095 & 44.699$\pm$0.005 & 4533.6$\pm$0.8 & 20.3$\pm$1.4 & 3.00$\pm$0.52 & 7.94$\pm$0.06 \\
J103837-2 & 56427.9 & 1.01 & -1.24$\pm$0.06 & 2.515$\pm$0.073 & 44.320$\pm$0.008 & 4534.8$\pm$0.7 & 18.2$\pm$1.5 & 2.37$\pm$0.55 & 7.61$\pm$0.07 \\
J103837-3 & 57133.9 & 0.96 & -1.38$\pm$0.01 & 5.227$\pm$0.037 & 44.638$\pm$0.002 & 4537.4$\pm$0.5 & 18.3$\pm$0.9 & 2.01$\pm$0.29 & 7.81$\pm$0.04 \\
		\hline
	\end{tabular}
\end{table*}

\begin{figure*}
	\includegraphics[width=2\columnwidth]{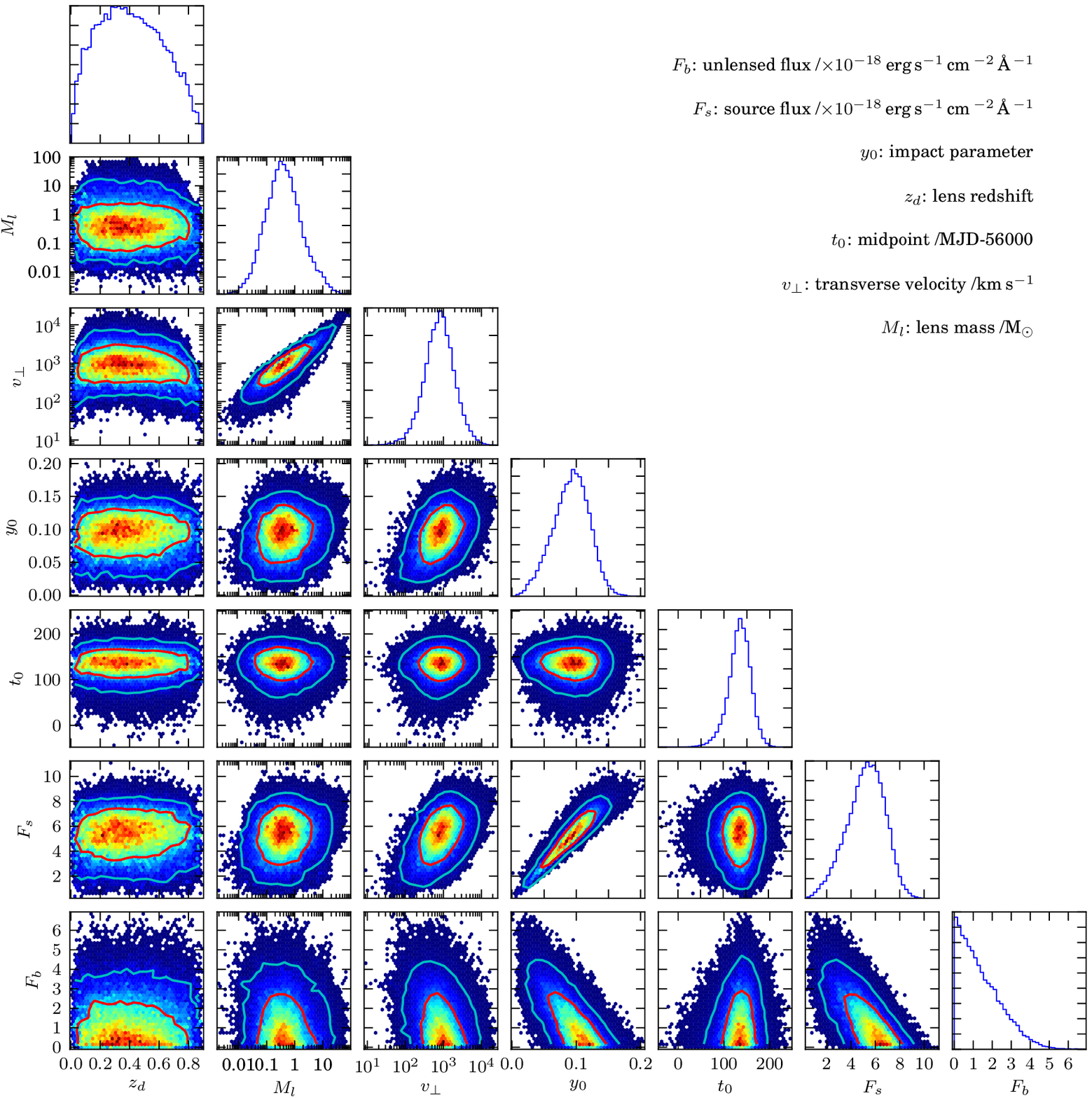}
    \caption{Corner plot from the MCMC analysis showing the one- and two-dimensional posterior probability distributions for J084305.}
    \label{fig:J084305_corner}
\end{figure*}

\begin{figure*}
	\includegraphics[width=2\columnwidth]{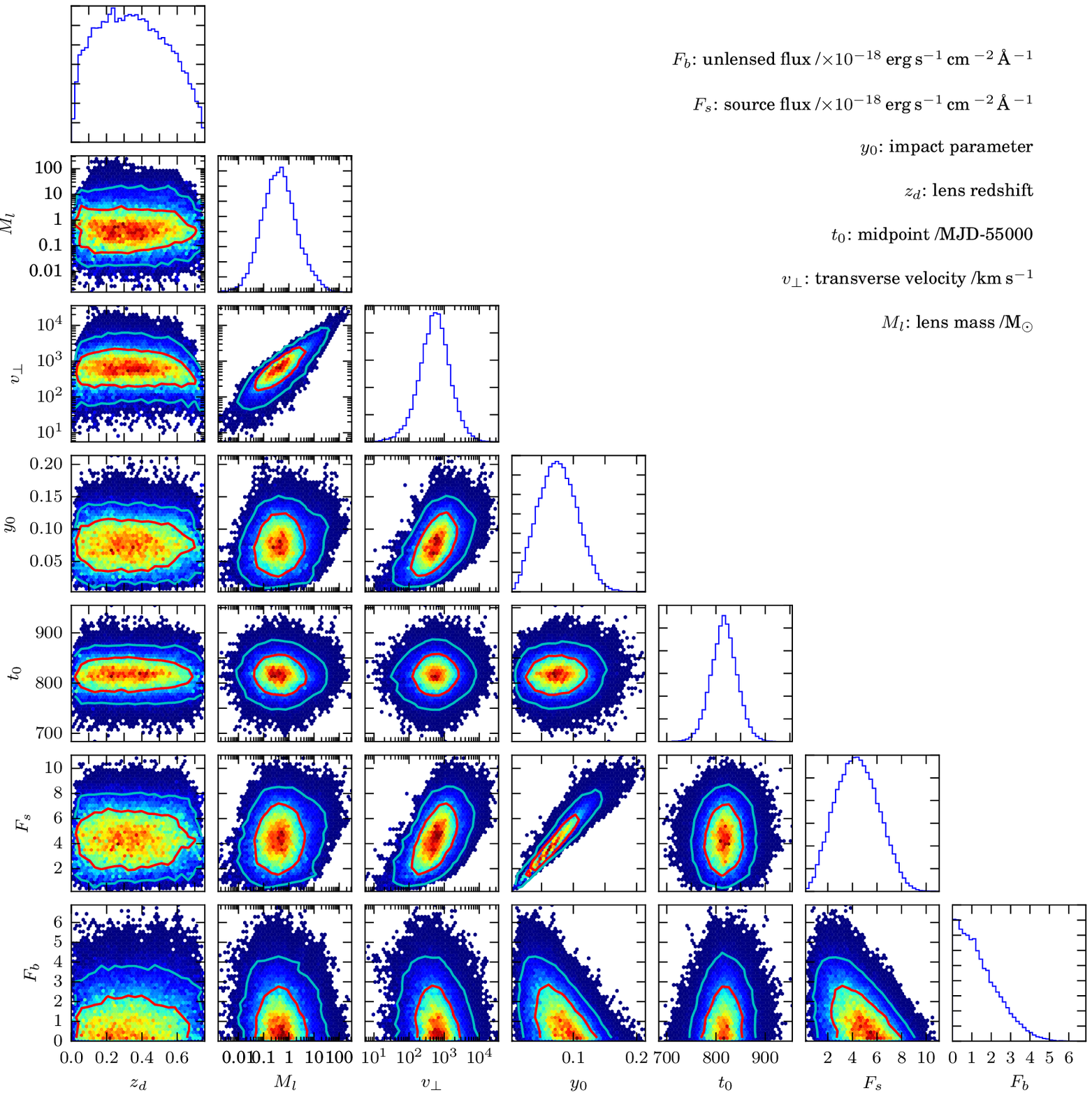}
    \caption{Corner plot from the MCMC analysis showing the one- and two-dimensional posterior probability distributions for J094511.}
    \label{fig:J094511_corner}
\end{figure*}

%%%%%%%%%%%%%%%%%%%%%%%%%%%%%%%%%%%%%%%%%%%%%%%%%%

% Don't change these lines
\bsp	% typesetting comment
\label{lastpage}
\end{document}